\documentclass[fleqn,usenatbib,useAMS]{mnras}
\pdfoutput=1

\usepackage{graphicx}    
\usepackage{amsmath}    
\usepackage{amssymb}    
\usepackage{multicol}        
\usepackage{bm}        
\usepackage{pdflscape}    
\usepackage{float}
\usepackage{placeins}

\usepackage[T1]{fontenc}
\usepackage{ae,aecompl}

\newcommand{\vsol}{\langle v_\mathrm{sol}^2 \rangle}
\newcommand{\vcomp}{\langle v_\mathrm{comp}^2 \rangle}

\title[Compression of magnetized gas subjected to radiative cooling]{\textbf{Molecular cloud formation by compression of magnetized turbulent gas subjected to radiative cooling}}

\author[Ankush Mandal, Christoph Federrath, Bastian K\"ortgen]{Ankush Mandal,$^{1,2}$\thanks{Contact e-mail: \href{mailto:ankushm@iucaa.in}{ankushm@iucaa.in}} Christoph Federrath,$^{1}$\thanks{Contact e-mail: \href{mailto:christoph.federrath@anu.edu.au}{christoph.federrath@anu.edu.au}} {Bastian K\"ortgen$^{3}$}\\
$^{1}$Research School of Astronomy $\&$ Astrophysics, The Australian National University, Canberra, ACT, Australia
\\
$^{2}$Inter-University Centre for Astronomy and Astrophysics, Pune-411007, India \\
$^{3}$Hamburger Sternwarte, Universit\"at Hamburg, Gojenbergsweg 112, D-21029 Hamburg, Germany}

\pubyear{2020}

\begin{document}
\label{firstpage}
\pagerange{\pageref{firstpage}--\pageref{lastpage}}
\maketitle

\begin{abstract}
Complex turbulent motions of magnetized gas are ubiquitous in the interstellar medium. The source of this turbulence, however, is still poorly understood. Previous work suggests that compression caused by supernova shockwaves, gravity, or cloud collisions, may drive the turbulence to some extent. In this work, we present three-dimensional (3D) magnetohydrodynamic (MHD) simulations of contraction in turbulent, magnetized clouds from the warm neutral medium (WNM) of the ISM to the formation of cold dense molecular clouds, including radiative heating and cooling. We study different contraction rates and find that observed molecular cloud properties, such as the temperature, density, Mach number, and magnetic field strength, and their respective scaling relations, are best reproduced when the contraction rate equals the turbulent turnover rate. In contrast, if the contraction rate is significantly larger (smaller) than the turnover rate, the compression drives too much (too little) turbulence, producing unrealistic cloud properties. We find that the density probability distribution function evolves from a double log-normal representing the two-phase ISM, to a skewed, single log-normal in the dense, cold phase. For purely hydrodynamical simulations, we find that the effective driving parameter of contracting cloud turbulence is natural to mildly compressive (\mbox{$b\sim0.4$--$0.5$}), while for MHD turbulence, we find \mbox{$b\sim0.3$--$0.4$}, i.e., solenoidal to naturally mixed. Overall, the physical properties of the simulated clouds that contract at a rate equal to the turbulent turnover rate, indicate that large-scale contraction may explain the origin and evolution of turbulence in the ISM.
\end{abstract}

\begin{keywords}
molecular cloud --ISM, magnetohydrodynamics -- turbulence 
\end{keywords}



\begingroup
\let\clearpage\relax
\endgroup
\newpage

\section{Introduction}
Molecular clouds (MCs) -- the birthplace of the stars -- have been a matter of interest for the last few decades. Extensive studies about the interstellar medium (ISM) and giant molecular clouds (GMC) have established that the gases in the ISM and MCs are highly magnetized and supersonically turbulent in nature. The star formation rate (SFR) in MCs is directly correlated with the physical properties of the clouds. For example, it is a complex competition between supersonic turbulence and self-gravity along with the column density, magnetic field, radiation, and thermal pressure that determines when and where stars form inside the clouds \citep{Low2004, Larson2005, McKee2007}. However, observations show that the rate of the formation of stars is much slower than that expected if the clouds were forming stars at a free-fall rate \citep{Zuckerman, Wong, Gao}. Thus, this indicates that there are physical processes that oppose the gravitational free-fall. The current understanding is that supersonic turbulence plays a crucial role in opposing the fast gravitational collapse \citep{Semadeni, Low2004, Federrath2012,Padoan2014,Krumholz2019}. However, it has been established that un-driven supersonic turbulence decays quickly, on a time scale comparable to the turnover time of the largest eddies \citep{Low1998, Padoan1999}. This means that turbulence must be driven by some physical mechanism \citep{Federrath2017}.

Here we study the maintenance and dissipation of turbulence in MCs in the context of star formation theories. Despite its importance and inevitability for star formation, the origin and evolution of the interstellar medium from the warm atomic phase to the cold, dense molecular clouds is still poorly understood. There are a number of proposed models that act as a source of driving of the turbulence, which include protostellar outflows \citep{Li2006,Wang2010,Federrath2014a}, feedback from massive stars such as expanding H\,\textsc{II} regions \citep{Matzner2002,Krumholz2006,Goldbaum2011}, energy injection from ongoing accretion \citep{Hennebelle2010,Vazquez2010,Lee2016}, gravitational contraction on small scales \citep{Federrathetal2011a,Sur2012} or supernova feedback \citep{Padoan2016a,Padoan2016b,Pan2016,Kortgen2016}. 

On the other hand, large-scale contraction of molecular clouds has gained some attention recently in the list of driving agents of the turbulence. Recent numerical and observational studies \citep{Ntormousi2011,Tremblin2014,Dawson2015} show that molecular clouds may be formed at the stagnation point between two expanding superbubbles due to the compression induced by the expansion, even without self-gravity. This motivates us to consider a large-scale global compression (not necessarily due to self-gravity) as a potential agent of driving the turbulence and a mechanism for molecular cloud formation as the global compression has the ability to pump energy into the turbulence to slow down the collapse \citep{Kortgen2017,Birnboim2018}.

Large-scale contraction in a turbulent medium has been studied for non-magnetized turbulence by \citet{Robertson_2012}, where they considered the equation of state to be isothermal and the compression can inject energy in a way that they have described as adiabatic heating. \citet{Birnboim2018} studied the same phenomenon, but for isothermal magnetohydrodynamic (MHD) turbulence. In this model, when the gas gets compressed, the velocity increases due to $P-V$ work (there $P$ and $V$ are the pressure and volume of the cloud) against the kinetic pressure (pressure that is generated by the kinetic motions of the particle hitting and rebounding from the surface). Due to compression, the eddy turnover timescale ($\tau \sim L/v$) decreases, and as a result, the dissipation rate increases. Thus, depending on the balance between compression timescale (here parameterised by a negative `Hubble' parameter, $H=\dot{a}/a$, where $a$ is the time-independent scale factor; see details in the method section) and the dissipation timescale ($\tau = 1/\omega = aL/(2v)$, where $\omega$, $L$ and $v$ are the turnover frequency, cloud size, and velocity dispersion, respectively), the turbulence can get amplified or dissipates away. 

However, \citet{Robertson_2012} and \citet{Birnboim2018} did not include the effects of radiative heating and cooling, which are crucial for the transition from the atomic to the molecular phase of interstellar clouds. Gas inside the MCs usually radiates its internal energy (radiative cooling) or absorbs energy from the incident radiation (radiative heating) through different complex mechanisms, and the cooling or heating rate depends on various physical parameters that have been studied extensively \citep{Cox, Raymond, Shull, Sutherland}. When the magnetized gas is subjected to rapid radiative cooling, the result is a highly supersonic flow (as the turbulent sonic Mach number is proportional to the inverse of temperature). Thus the effect of cooling has the potential to alter the dynamics of the cloud \citep{Koyama2002, Vazquez2007}. Moreover, a model for molecular cloud evolution is not complete, if it only predicts the source of the driving of turbulence, but not the formation of the cloud itself. A successful model would also reproduce the physical properties of the clouds that have been measured through different observational techniques \cite[see][for a detailed overview and references therein]{Heyer}.

From various theoretical models and observational surveys, it has been established that MCs are highly supersonic and magnetized with Mach number ($\mathcal{M})\sim 5-20$, temperature ($T)\sim10-50\,\text{K}$, density range, $n\sim 10^2-10^5\,\text{cm}^{-3}$  \citep{Wilson1997,Hughes}. \citet{Larson1981} first pointed out that there is a strong correlation between the velocity dispersion ($\sigma_v$) and the size of the cloud ($L$), and established a scaling relation ($L-\sigma_v$ scaling relation) in the form of a power law, $\sigma_v \propto \ell^{0.5}$ ($\ell$ is the cloud size in the unit of pc), that has been verified observationally \citep{Crutcher1999,S87,Ossenkopf2002,Heyer2004,Roman-Duval2011} . In addition, the magnetic field strength ($B$) also shows a correlation with the number density of the cloud in the high-density regime, $n\gtrsim 10^3\,\text{cm}^{-3}$ \citep{Crutcher2012}.

In this work, we aim to extend the works by \citet{Robertson_2012} and \citet{Birnboim2018}. In particular, we seek to determine whether the effects of heating and cooling, which were not included in \citet{Robertson_2012} and \citet{Birnboim2018}, can change the dynamics and structure of the MCs. To this end, we run three-dimensional (3D) hydrodynamic (HD) and MHD simulations including equilibrium heating and cooling. We consider different compression rates with respect to the eddy turnover rate to figure out the dependence of physical properties of the molecular clouds on the global compression rate, and to determine which contraction model is most favorable in the context of molecular cloud formation and evolution by comparing the results from simulations with theoretical models and observational predictions. We organize the paper in the following way. In \S\ref{Sec 2} we discuss the detailed methodology of our simulations, the physics of equilibrium cooling and its implementation. In \S~\ref{Sec 3}, we report the results from simulations. In \S~\ref{Sec 4} we briefly describe the limitations of our study, and in \S~\ref{Sec 5} we summarize our conclusions.

\section{Simulation methods}
\label{Sec 2}
\subsection{The FLASH code}
\label{Sec 2.1}
We use the modified version of the grid-based code FLASH \citep{Flash} to solve the three-dimensional (3D), compressible, ideal MHD equations,
\begin{gather}
    \frac{\partial}{\partial t}\rho + \boldsymbol{\nabla}\cdot(\rho\boldsymbol{v})=0,\label{static mass conservation} \\
    \frac{\partial}{\partial t}(\rho\boldsymbol{v})+\boldsymbol{\nabla}\cdot\big(\rho\boldsymbol{v}\otimes\boldsymbol{v}-\frac{1}{4\pi}\boldsymbol{B}\otimes\boldsymbol{B}\big)+\boldsymbol{\nabla} P_{\text{tot}}=0, \label{static momentum} \\
    \frac{\partial}{\partial t}e+\boldsymbol{\nabla}\cdot\Big[(e+P_{\text{tot}})\boldsymbol{v}-\frac{1}{4\pi}(\boldsymbol{B}\cdot\boldsymbol{v})\boldsymbol{B}\Big]=\frac{1}{\rho}\Big[\frac{\rho}{\mu m_\mathrm{H}}\Gamma-\Big(\frac{\rho}{\mu m_\mathrm{H}}\Big)^2\Lambda(T)\Big], \label{static energy} \\
    \frac{\partial}{\partial t}\boldsymbol{B}-\boldsymbol{\nabla}\times(\boldsymbol{v}\times\boldsymbol{B})=0, \hspace{0.2 cm} \boldsymbol{\nabla}\cdot\boldsymbol{B}=0. \label{static magneticfield}
\end{gather}
Here, $\rho$, $\boldsymbol{v}$, $P_{\text{tot}}=P_{\text{th}}+(1/8\pi)|{\boldsymbol{B}}|^2$, $\boldsymbol{B}$, $\mu$ and $e=\rho\epsilon_{\text{int}}+(1/2)\rho|\boldsymbol{v}|^2+(1/8\pi)|\boldsymbol{B}|^2$ denote the gas density, velocity, total pressure (including thermal and magnetic), magnetic field, mean molecular weight of the particles and total energy density (internal, kinetic and magnetic). For simplicity, we assume $\mu=1$ throughout this study, which does not affect the main conclusions of this study (see Sec.~\ref{Sec 4}). The MHD equations are closed by the ideal gas equation of state,
\begin{equation}
    P_{\text{th}}=(\gamma-1)\rho\epsilon_{\text{int}}, \label{EoS}
\end{equation}
where we assume $\gamma=5/3$ throughout.
The energy equation also includes the heating ($\Gamma$) and cooling ($\Lambda$) terms, which we discuss in Sec.~\ref{sec 2.3}.
To solve the system of MHD equations (\ref{static mass conservation}-\ref{static magneticfield}), we use the robust HLL3R Riemann scheme by \citet{Waagan}, based on previous developments in applied mathematics, to maintain positive density and pressure.

\begin{table*}
    \caption{List of different simulation parameters at the beginning of the contraction}
    \label{Table 1}
    \begin{tabular*}{\textwidth}{c @{\extracolsep{\fill}} cccccccc}
    \hline
    \hline
    Model & $H\,(\text{s}^{-1})$ & $\tau=1/\omega$ (s) & $\omega/H$ & $B$ ($\mu \text{G}$) & $\sigma_v\,(\text{km/s)}$ & $\mathcal{M}$ & $T\,(\text{K})$ & $N_{\text{res}}^3$  \\
    \hline
    HD-Slow & $-3.241\times10^{-16}$ & $3.08\times10^{14}$ &-0.1 & 0 & 11.57 & 1.85 & $4.67\times 10^4$ & $(512)^3$ \\
    HD-Medium & $-3.241\times10^{-15}$ & $3.08\times10^{14}$ & -1.0 & 0 & 11.57 & 1.85 & $4.76\times10^4$ & $(512)^3$ \\
    HD-Fast & $-3.241\times10^{-14}$ & $3.08\times10^{14}$ & -10.0 & 0 & 11.52 & 1.84 & $4.78\times10^4$ & $(512)^3$ \\
    \hline
    MHD-Slow & $-3.241\times10^{-16}$ & $3.08\times10^{14}$ & -0.1 & 1.82 & 11.12 & 1.75 & $4.97\times10^4$ & $(512)^3$ \\
    MHD-Medium & $-3.241\times10^{-15}$ & $3.08\times10^{14}$ & -1.0 & 1.82 & 11.12 & 1.75 & $4.97\times10^4$ & $(512)^3$ \\
    MHD-Fast & $-3.241\times10^{-14}$ & $3.08\times10^{14}$ & -10.0 & 1.75 & 11.22 & 1.75 & $4.87\times10^4$ & $(512)^3$ \\
    \hline
    \end{tabular*}
\end{table*}

\subsection{MHD equations in a contracting reference frame}
\label{Sec 2.2}
Although the default hydrodynamic scheme in FLASH is written for a static frame of reference, one can use the cosmology module in FLASH to solve the MHD equations in an expanding or contracting frame of reference. In order to do this, we change the MHD equations from the physical coordinate system to the co-moving coordinate system, where additional terms appear due to contraction or expansion. All calculations are assumed to take place in co-moving coordinates $\boldsymbol{x}=\boldsymbol{r}/a$, where $\boldsymbol{r}$ is the physical position vector and $\boldsymbol{x}$ is the co-moving position vector. $a(t)$ is the dimensionless scale factor, which depends on time. The transformation of time and space derivatives in co-moving coordinates is related to the proper coordinates by  $(\partial/\partial t)_{\boldsymbol{x}}$ = $(\partial/\partial t)_{\boldsymbol{r}}+H\boldsymbol{r}\cdot\boldsymbol{\nabla_{\boldsymbol{r}}}$ and $\boldsymbol{\nabla_x}$ = $a\boldsymbol{\nabla_r}$, where the Hubble constant ($H$) is defined as $H=\dot{a}/a$. The physical velocity is $\Tilde{\boldsymbol{v}}=H\boldsymbol{r}+a\dot{\boldsymbol{x}}$, where the first term is the Hubble flow and the second term is called peculiar velocity, i.e. the velocity in the co-moving frame of reference.

The hydrodynamic quantities in the physical (with tilde) and co-moving (without tilde) coordinate system are related by the following equations,
\begin{gather}
    \rho=a^3\Tilde{\rho}, \label{density scaling} \\
    P_{\text{tot}}=a\Tilde{P}_{\text{tot}}, \label{pressure scaling} \\
    e=a\Tilde{e}, \label{energy scaling} \\
    \epsilon_{\text{int}}=a^{-2}\Tilde{\epsilon}_{\text{int}}, \label{int energy scaling} \\
    \boldsymbol{B}=a^{1/2}\Tilde{\boldsymbol{B}} \label{magnetic scaling},
\end{gather}
The MHD equations in co-moving coordinates can be determined using a definition of a time and space derivative along with prior hydrodynamic quantities, which read,
\begin{gather}
    \frac{\partial}{\partial t}\rho + \boldsymbol{\nabla}\cdot(\rho\boldsymbol{v})=0, \label{co-moving mass conservation} \\
    \frac{\partial}{\partial t}(\rho\boldsymbol{v})+\boldsymbol{\nabla}\cdot\big(\rho\boldsymbol{v}\otimes\boldsymbol{v}-\frac{1}{4\pi}\boldsymbol{B}\otimes\boldsymbol{B}\big)+\boldsymbol{\nabla} P_{\text{tot}}=-2H\rho\boldsymbol{v}, \label{co-moving momentum}
\end{gather}
\begin{multline}
\frac{\partial}{\partial t}e+\boldsymbol{\nabla}\cdot\Big[(e+P_{\text{tot}})\boldsymbol{v}-\frac{1}{4\pi}(\boldsymbol{B}\cdot\boldsymbol{v})\boldsymbol{B}\Big]= \\
-H[(3\gamma-1)\rho\epsilon_{\text{int}}+2\rho\boldsymbol{v}\cdot\boldsymbol{v}]+\frac{1}{\rho}\Big[\frac{\rho}{\mu m_{\text{H}}}\Gamma-a^{-3}\Big(\frac{\rho}{\mu m_{\text{H}}}\Big)^2\Lambda(T)\Big], \label{co-moving energy}
\end{multline}
\begin{gather}
    \frac{\partial}{\partial t}\boldsymbol{B}-\boldsymbol{\nabla}\times(\boldsymbol{v}\times\boldsymbol{B})=-\frac{3}{2}H\boldsymbol{B}, \hspace{0.2 cm} \boldsymbol{\nabla}\cdot\boldsymbol{B}=0, \label{co-moving magnetic field}
\end{gather}
where $\partial/\partial t \equiv (\partial/\partial t)_{\boldsymbol{x}}$ and $\boldsymbol{\nabla}\equiv\boldsymbol{\nabla_{x}}$ are the derivatives in the co-moving frame. We use operator splitting to account for the Hubble source terms, where the co-moving hydrodynamic variables are modified in each time step to account for the expansion/contraction \citep{Birnboim2018}.

\subsection{Radiative heating and cooling}
\label{sec 2.3}
The previous studies by \citet{Robertson_2012} and \citet{Birnboim2018} used an isothermal equation of state. However, in the real ISM, gas can absorb or emit radiation depending on the quantum state and composition of the gas. There are various mechanisms that can heat or cool: photoelectric heating from small grains and polycyclic aromatic hydrocarbons, heating and ionization from cosmic rays and X-rays, $\text{H}_2$ formation and destruction, atomic line cooling from hydrogen, etc.~\citep{Sutherland}. As a result, the temperature of the cloud varies, depending on the balance between these various heating and cooling processes. The heating or cooling rate depends on the temperature of the gas cloud, which again depends on the density ($\rho$). As for static turbulence, the mean density remains constant, the temperature does not vary that much, which means the cooling rate is almost constant throughout the evolution. But, for compressing turbulence, the mean gas density increases with time, and hence, temperature varies a lot. As a result, the heating or cooling rate varies, which has a profound effect on the evolution of the turbulence.

Here we use tabulated values for $\Gamma$ and $\Lambda$ developed by \citet{Koyama2002} and \citet{Vazquez2007}, based on a constant heating rate,
\begin{equation}
    \Gamma=2\times10^{-26}\,\text{erg s$^{-1}$} \label{heating rate},
\end{equation}
and a cooling rate based on the following equation,
\begin{multline}
    \frac{\Lambda(T)}{\Gamma}=10^7\text{exp}\Big(\frac{-1.184\times10^5}{T+1000}\Big) \\ +1.4\times10^{-2}\sqrt{T}\text{exp}\Big(\frac{-92}{T}\Big)\,\text{cm$^3$}, \label{cooling rate}
\end{multline}
where the temperature $T$ is in units of Kelvin. These functions are the fits to heating ($\Gamma$) and cooling ($\Lambda$) due to various processes mentioned above. The thermal equilibrium condition is given by,
\begin{equation}
    n\Gamma=n^2\Lambda, \label{equilibrium condistion}
\end{equation}
where $n=\rho/m_{\text{H}}$ is the number density with $\mu=1$ for this study.

In hydrodynamic simulations, we generally apply cooling by considering the cooling rate in the Courant condition to limit the simulation time step. The densities gradually increase in this problem of gas compression, which means a very small time step can occur as the cooling rate increases. To avoid this problem, we treat cooling as a source term in operator splitting, and following each hydrodynamic step, the internal energy is adjusted. Consider $T_{\text{eq}}$ and $e_{\text{eq}}$ are the equilibrium temperature and internal energy, and the time required to radiate or absorb excess thermal energy is
\begin{equation}
    \tau_{\Lambda}=\left|{\frac{e-e_\mathrm{eq}}{n^2\Lambda-n\Gamma}}\right| \label{cooling time scale}
\end{equation}
Let $\epsilon$ be the excess energy. The rate of change of energy with time is directly proportional to the instantaneous energy (Newton's cooling law). Thus we have,
\begin{equation}
    \frac{d\epsilon}{dt}=-\frac{\epsilon}{\tau_{\text{ch}}} \label{Newton's cooling}
\end{equation}
Here $\tau_{\text{ch}}$ is the characteristic cooling time-scale. Now, if the excess energy after time $t$ is $\epsilon_1$, then,
\begin{equation}
    \epsilon_1=\epsilon\,\mathrm{exp}(-t/\tau_{\mathrm{ch}})
\end{equation}
In this case the initial excess internal energy is $\Delta e = e-e_{\text{eq}}$. So, after a time step $dt$ the excess internal energy will be $\Delta e'=(e-e_{\text{eq}})\,\mathrm{exp}(-dt/\tau_{\Lambda})$. Then we compute the new internal energy $e'$, after a time step $dt$, as
\begin{equation}
    e'=e_{\text{eq}}+(e-e_{\text{eq}})\,\text{exp}\left(\frac{-dt}{\tau_{\Lambda}}\right). \label{int energy modification}
\end{equation}
From this equation, we see that if the gas is undergoing rapid cooling (or heating), $\tau_{\Lambda}\ll dt$ and $\mathrm{exp}(-dt/\tau_{\Lambda})\rightarrow 0$, such that the gas reaches thermal equilibrium very quickly. On the other hand, if the cooling (or heating) rate is very slow, then $\tau_{\Lambda} \gg dt$ and Eq.~\eqref{int energy modification} reduces to
\begin{equation}
    e'=e_{\text{eq}}-dt\,(n^2\Lambda-n\Gamma).
\end{equation}

\subsection{Initial driving of turbulence to generate initial conditions}
\label{Sec 2.4}
As we are experimenting with MHD turbulence statistics in a contracting reference frame, we need a fully-developed turbulent field as the initial conditions of the contraction phase. To do that we first drive the turbulence for five eddy turnover times, $\tau=1/\omega=L/(2\sigma_v$), on a static background ($a=1$). Here we consider a box size of $L=200\,\text{pc}$, i.e., covering a large portion of the warm neutral ISM with a uniform density of $1\,\mathrm{cm}^{-3}$. We drive turbulence to reach a velocity dispersion $\sigma_v=10\,\text{km/s}$, typical of the velocity dispersion of the Milky Way on large scales (of order the disc scale height).

The turbulence is driven by applying $\rho\boldsymbol{F}$ as a source term in the momentum equation~\eqref{static momentum}. In developing the turbulence acceleration field $\boldsymbol{F}$, we use the stochastic Ornstein-Uhlenbeck (OU) method \citep{Eswaran,Schmidt,Price}. \citet{Federrathetal2010} developed the code, which is accessible in the public version of FLASH. Turbulent stirring from larger scales is correlated on timescales related to the lifetime of an eddy on the scale of the simulation domain. The OU process is a well-defined stochastic process with finite auto-correlation timescale. In our periodic simulation box of side length $L$, it produces a smoothly varying spatial and temporal driving pattern on the largest scales ($L/2)$. The driving process is carried out in Fourier space, and the acceleration field $\boldsymbol{F}$ is set to inject most of the energy into the lowest wave numbers, $1<|\boldsymbol{k}|L/2\pi<3$. The spectral shape of the driving field we choose is paraboloid, i.e ,the peak energy injection is on scale $L/2$, and falls off as a parabola for smaller and higher wave number, so that the energy injection at $\boldsymbol{k}=2\pi/L$ and $\boldsymbol{k}=6\pi/L$ is identically zero \citep{Federrathetal2010,Federrath2013,federrath2016,Birnboim2018}.

Depending on the physical interests, we can build the driving field either purely solenoidal ($\boldsymbol{\nabla}\cdot\boldsymbol{F}=0$) or compressive ($\boldsymbol{\nabla}\times\boldsymbol{F}=0$) or a blended field with fractional solenoidal and compressive modes. For separating the driving field into the solenoidal and compressive components, we use the Helmholtz decomposition in Fourier space. For simplicity, we here use only solenoidal driving to develop the turbulence fully before starting the contraction phase ($a<1$).
\begin{figure}
    \centering
    \includegraphics[width=0.99\linewidth]{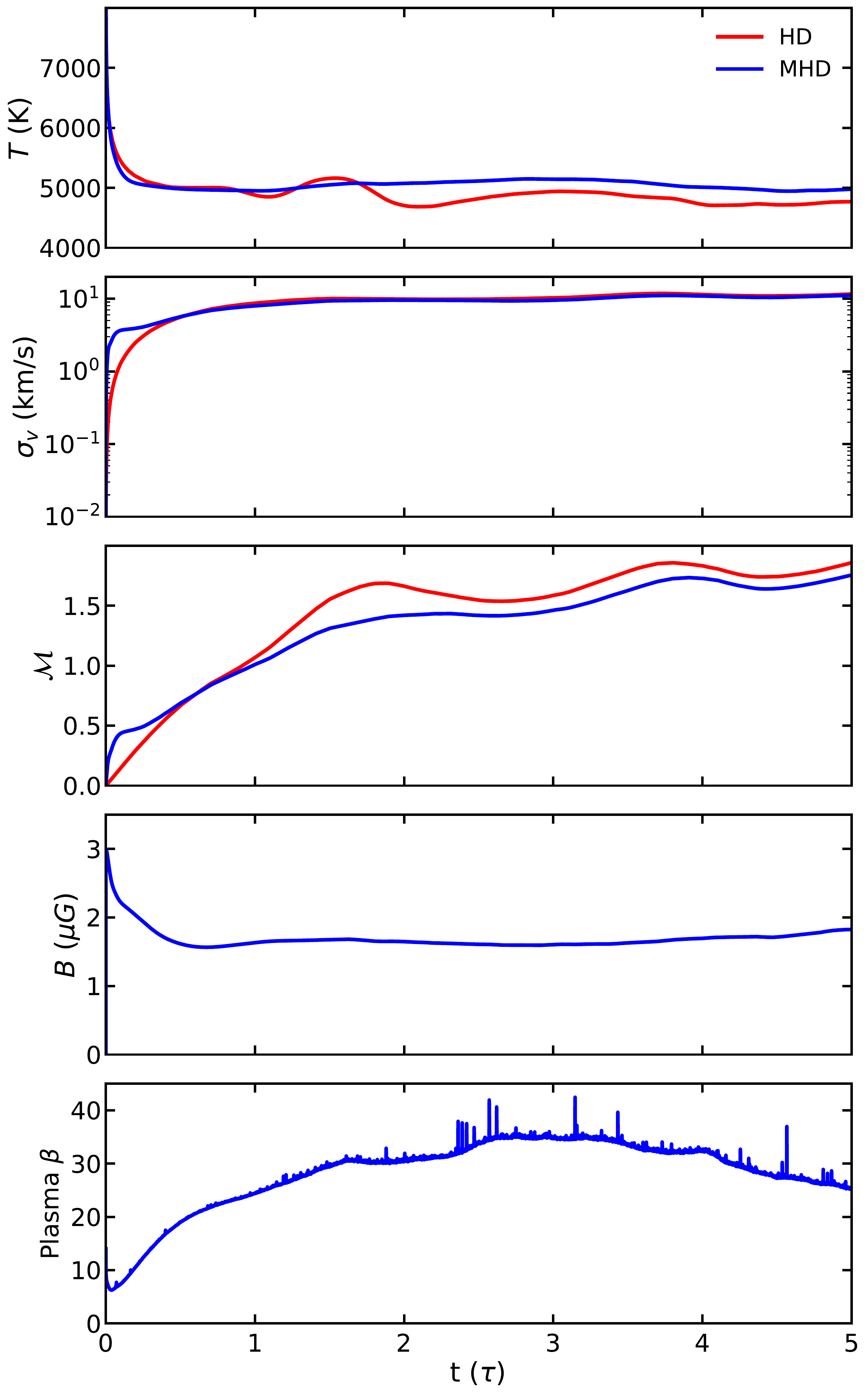}
    \caption{Initial driving phase to establish fully-developed turbulence. We take the state at after five eddy turnover times ($t=5\,\tau$) to serve as the initial condition for the contraction phase. The red line corresponds to the purely hydrodynamic case (HD) and the blue line is for the MHD case.}
    \label{Figure 1}
\end{figure}

\subsection{Construction of initial turbulent magnetic field}
\label{Sec 2.5}
The interstellar medium is magnetized. Thus, in order to simulate MHD turbulence, we have to set the initial magnetic field. In our study, we chose the magnetic field structure completely random, that is to say, the magnetic field is fully turbulent. In order to construct a fully turbulent magnetic field we use a method that has been considered in the studies of \citet{Gerrard} and \citet{Birnboim2018}. In this technique, we generate the initial conditions so that all field vectors are randomly oriented, instead of driving the turbulence in the field. We use a Kazantsev power spectrum with an exponent 3/2 to decompose the turbulent field in Fourier space \citep{Brandenburg,federrath2016}. We restrict the wave vectors in the range $2<|\boldsymbol{k}|/2\pi<20$. The Kazantsev spectrum comes from turbulent dynamo amplification \citep{Kazantsev, Federrathetal2011a} as field amplification works on the small-scale seeds of the magnetic field \citep{Brandenburg,Schober,Schleicher}.

The typical values of the large-scale magnetic field in interstellar clouds in the spiral arms of the Milky Way and also in nearby galaxies are about $3-10\,\mu\mathrm{G}$ \citep{Beck2015,Han}. Therefore, for our MHD simulations we initially set $B$ = $3\,\mu\text{G}$, and after $t=5\,\tau$ it has slightly relaxed to $B=1.8\mu\mathrm{G}$. Thus, the initial magnetic field is slightly weaker than the observed field, but the MHD simulations nevertheless provide us with at least a reasonable qualitative measurement of MHD effects during cloud contraction.


\subsection{Initial conditions, contraction parameters, and list of simulations}
\label{Sec 2.6}
All the simulations started from a uniform density $\rho_0=1.67\times10^{-24}\,\text{g-cm}^{-3}$ and zero velocities. The simulation box size is initially $L=200\,\text{pc}$. Then the turbulence was driven to five eddy turnover times on a static background to establish fully developed turbulence. After that, the driving module has been disabled and the cosmology module has been activated and the evolution followed based on the cosmological factor $a(t)$. The scale factor $a(t)$ of the compression is solely determined by the Hubble parameter $(H)$: $a(t)=\text{exp}[H(t-t_0)]$, where $t\geq t_0$ ($t_0$ is the contraction start time). In this phase $a(t)<1$ (Hubble parameter $H$ is negative) as the box started contracting, and the dynamics are determined by the contraction.
\begin{figure*}
  \centering
  \includegraphics[width=0.99\linewidth]{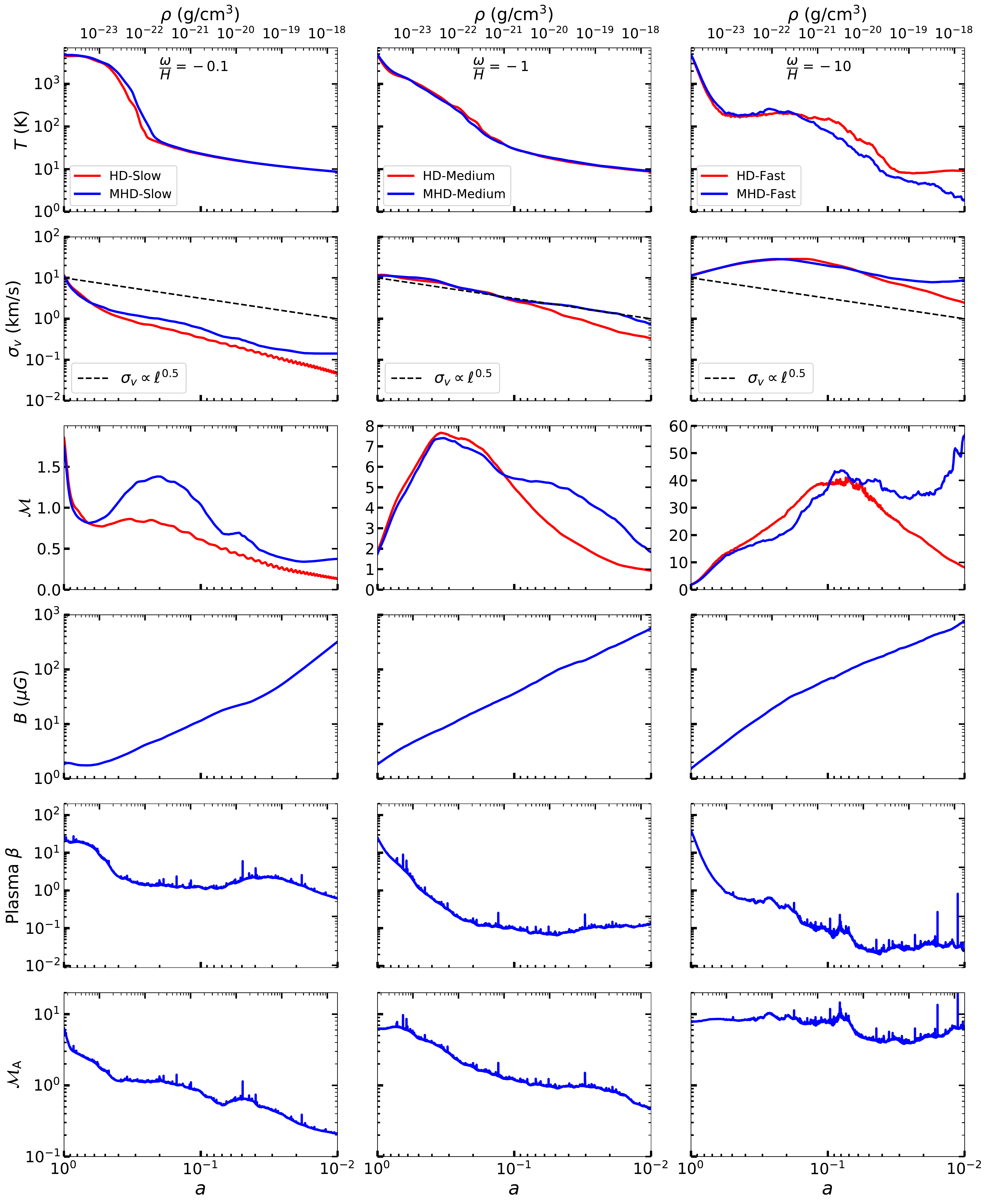}
  \caption{Integrated quantities as a function of scale factor $a$ (bottom axis) and mean density (bottom axis). The different column show (from the left to right) slow($\omega/H=-0.1$), medium ($\omega/H=-1$) and fast ($\omega/H=-10)$) compression. The red lines show the purely hydrodynamic case, and blue lines are for the magneto-hydrodynamic case. In the velocity dispersion panel (second row) we have plotted the linewidth-Size scaling relation (Eq.~\ref{velocity dispersion scaling}). Please note the different range of sonic Mach numbers (third row) for the three compression rates. }
  \label{Figure 2}
\end{figure*}

In order to study the dependence of the dynamics of turbulent gas on the contraction rate, we chose three values of the Hubble parameter ($H$): 1.~slow compression (the contraction time scale is $10$ times longer than the eddy turnover time, $\omega/H=-0.1$), 2.~medium compression (contraction time scale is equal to eddy turnover time, $\omega/H=-1$), and 3.~fast compression (contraction time scale is $10$ times shorter than eddy turnover time, $\omega/H=-10$). Table~\ref{Table 1} provides the list of all simulations and initial conditions that have been used for the contraction phase. Fig.~\ref{Figure 1} shows the evolution of important integrated quantities during the driving phase.

In reality, the large-scale global compression may be induced by various agents, such as supernova explosions, self-gravity, dynamical shocks (from e.g., spiral-arm dynamics), compression induced by ionization fronts, cloud-cloud collisions and many more. Thus, it is not clear what should be the combined compression rate and exact function form caused by all of these mechanisms. Thus, we focus on a simple model of compression, where the compression rate is constant and given in units of the turbulent turnover rate, and we systematically vary this contraction rate over different simulations to study the effect of different contraction rates on the dynamics and cloud properties produced.

Previously, \citet{Robertson_2012} and \citet{Birnboim2018} have discussed cases with different contraction rates for pure HD and MHD simulations for isothermal turbulence. Here we are mainly focused on the effect of radiative cooling on contracting background with different contraction rates for pure HD and MHD turbulence in the ISM.

\section{RESULTS AND DISCUSSIONS}
\label{Sec 3}
\subsection{Evolution of integrated quantities }
\label{Sec 3.1}
In this section we present the main results obtained from our numerical simulations of contracting interstellar clouds that were initialized with the final state shown in Fig.~\ref{Figure 1}. The evolution of important integrated quantities during the contraction phase is shown in Fig.~\ref{Figure 2} as a function of scale factor ($a$). The domain size $L(a)=aL(a=1)$, such that the cloud has contracted from $200\,\mathrm{pc}$ at $a=1$ down to $2\,\mathrm{pc}$ at $a=0.01$. All the quantities are plotted with decreasing $a$ (on the bottom x-axis) and increasing mean density (on the top x-axis). As the scale factor ($a$) is exponential in time $a(t)=\text{exp}[H(t-t_0)]$ and the $a$-axis is in logarithmic scale, it shows the evolution proportional to time. The quantities shown were integrated over the whole volume of the simulation domain. We divide our simulation results in three parts. The first column in Fig.~\ref{Figure 2} presents the results for slow compression ($\omega/H=-0.1$). The second and third column correspond to medium ($\omega/H=-1$) and fast ($\omega/H=-10$) compression, respectively. For each case we consider two types of simulations: 1.~a purely hydrodynamic simulation (HD), and 2.~a magneto-hydrodynamic simulation without guide field (MHD). 

\subsubsection{Evolution of temperature}
\label{Sec 3.1.1}
The first row of Fig.~\ref{Figure 2} shows the temperature evolution. The initial temperature for all the simulations is approximately the same ($\sim 5000\,\text{K}$, the equilibrium temperature of gas with mean density $\sim1\,\mathrm{cm}^{-3}$). As the compression starts the temperature drops. The behavior of the temperature evolution is similar for each model, because it is primarily determined by the evolution of the mean density, which controls the cooling rate. However, the initial slope of the temperature curve is different for different models. For fast compression the number density grows faster, causing a higher cooling rate and a faster drop in temperature until the density has reached $\sim10^{-23}\,\mathrm{g-cm}^{-3}$. Another point to notice is that the presence of a magnetic field does not change the behavior much, as the HD and MHD simulations for each model follow almost the same temperature evolution. This is expected as 1) the field is weak and does not affect much the density, 2) We do not use a proper heating/cooling via a chemical network, where the species abundances might be affected by the magnetic field. Initially, when the temperature is about $\sim 5000\,\text{K}$, the cloud is mostly in the warm atomic phase. As the thermal energy of the particles drops below (due to cooling) the binding energy ($T\sim 150\,\text{K}$) of H$_2$ molecules, the atomic hydrogen undergoes a phase transition to form molecular hydrogen (H$_2$). From Fig.~\ref{Figure 2} we find this occurs at $a \sim 0.2$ for slow compression, $a\sim 0.3$ for medium compression and $a\sim 0.5$ for fast compression. Beyond this epoch, the fraction of molecular H$_2$ increases gradually and the gas becomes almost fully molecular. When the density $n\sim 10^3$--$10^6\,\text{cm}^{-3}$, the temperature drops to $\sim 50$--$10\,\text{K}$ and gets saturated around $\sim 10\,\text{K}$ beyond that. Observations in the Milky Way and extragalactic environments, such as the SMC and LMC, indicate a similar overall temperature dependence on density \citep{Wilson1997,Spitzer,Gratier,Hughes,Heyer,Katey2019}. However, there are details in the temperature evolution that clearly depend on the ratio of turbulent turnover rate and contraction rate, which are too subtle to distinguish in observations, based on the temperature evolution alone. Thus, we now turn to more statistics involving the velocity and magnetic field.

\subsubsection{Evolution of velocity dispersion ($\sigma_v$)}
\label{Sec 3.1.2}
An important distinction between different models is provided by the variation of velocity dispersion ($\sigma_v$). The second row of Fig.~\ref{Figure 2} presents the $\sigma_v$ evolution with scale factor. Starting from the fully-developed turbulence of $\sigma_v \sim 10\,\text{km/s}$, all the simulations start compression, which --depending on the contraction rate-- can drive or maintain turbulence to some level. For slow compression ($\omega/H=-0.1$), the compression timescale is longer than the eddy turnover timescale. As a result, the dissipation rate dominates over the compression rate and the compression cannot inject energy with enough power to drive the turbulence. Thus, $\sigma_v$ declines more steeply with $a$ than in the other cases. For fast compression ($\omega/H=-10$), $\sigma_v$ initially increases and after reaching a peak value it decays with time. This behaviour of increasing to decaying turbulence can be explained by the change of dissipation rate with $a$.
As the dissipation timescale is proportional to the largest eddy turnover time ($\omega \sim v/2aL$) \citep{Mac-Low1999,Robertson_2012,Birnboim2018} and $a$ decreases with time, the dissipation timescale decreases. After some point ($a\sim 0.2$) when $\omega/H$ becomes less than $-1$, the turbulence dissipation dominates.

For medium compression, the compression and dissipation rates are comparable and they remain so for a longer period of time, i.e., the contraction drives just enough turbulence to maintain a nearly constant decline of velocity dispersion with scale, very close to the observed scale dependence of $\sigma_v$. We discuss the scaling relation in details in Sec.~\ref{Sec 3.2}.

\subsubsection{Evolution of Mach number ($\mathcal{M}$)}
\label{ Sec 3.1.3}
The third row of Fig.~\ref{Figure 2} show the evolution of mass-weighted Mach number ($\mathcal{M} = \sigma_v / c_\mathrm{s}$) for different simulations. In the cold, dense molecular regime, there is no profound difference between the volume-weighted and mass-weighted Mach number. Here, we chose the mass-weighted Mach number, because it better represents the kinematics in the cold, dense phase, which is where the Mach number may be an important physical quantity to determine the star formation potential of clouds \citep{Federrath2012,Federrath2013Star,Salim2015,Sharda2018,Sharda2019,Beattie2019a,Beattie2019b}. The sound speed ($c_s$) directly depends on the temperature of the cloud ($c_s^2=\gamma k_BT/\mu m_\text{H}$, where $\gamma$ and $\mu$ are the adiabatic index and mean molecular weight of the gas particles, respectively). Initially, for all models, the turbulence is supersonic. Since contraction forces the temperature to drop, the sound speed ($c_s$) decreases with $a$, and $\mathcal{M}$ increases, which is a direct consequence of the joint evolution of temperature and velocity. For the medium and fast compression rate, the HD simulations show the same behavior, i.e $\mathcal{M}$ grows to a peak value and then decays. For slow compression, $\mathcal{M}$ decreases initially, and then grows to a peak value, followed by decay. This behavior is due to the different rates of change of temperature and velocity dispersion. The interesting point is to notice the behavior of $\mathcal{M}$ in the molecular cloud regime for different compression rates. The typical values of $\mathcal{M}$ in real molecular clouds are known to be supersonic with $\mathcal{M}\sim 5-20$ \citep{Crutcher1999,Schleicher}. From Fig.~\ref{Figure 2} we see that the Mach number in the molecular regime for slow compression is subsonic, which is too small. On the other hand, for fast compression, the Mach numbers exceed 30 in the MHD case, which is unusual for Milky Way conditions \footnote{For MHD turbulence, the presence of a magnetic field can change the situation considerably. In Fig.~\ref{Figure 2}, we see that the $\sigma_v$ as well as $\mathcal{M}$ stop decreasing and increases again after a sufficiently long period for MHD simulation in the case of fast compression. \citet{Birnboim2018} have shown that the value of $a$ where the transition from decaying to increasing turbulence happens depends on the presence of the guide field, and the saturation level of turbulent dynamo for isothermal compressive turbulence. The presence of a strong magnetic field changes the flow pattern to nearly dissipationless. An interesting point to note that although \citet{Birnboim2018} pointed out it for isothermal turbulence, the behavior does not change for the case of turbulence subjected to radiative cooling.}. Only the simulation with medium compression rate produces realistic Mach numbers of order $5-10$ in the molecular regime and a dependence on scale consistent with the observed velocity dispersion -- size relation.

\subsubsection{Magnetic field and plasma-$\beta$}
\label{Sec 3.1.4}
The fourth row of Fig.~\ref{Figure 2} shows the magnetic field evolution for the three MHD simulations with different compression rates. For all the simulations, the mean magnetic field ($|\boldsymbol{B}|$) starts from $|\boldsymbol{B}|\approx2\mu G$ after the initial driving phase (c.f., Fig.~\ref{Figure 1}) and starts growing due to the compression of field lines. Observations with different techniques like Zeeman splitting in H\textsc{I}, OH, CN absorption lines \citep{Crutcher1993,Crutcher1999,Falgarone}, and maser emission from dense molecular cloud cores \citep{Vlemmings,Watson2009} have shown the existence of magnetic fields in interstellar clouds. All of these studies show that for low density clouds ($n \lesssim 10^{3}\,\text{cm}^{-3}$), there is nearly no correlation between the magnetic field strength ($B$) and the density ($\rho$). However for dense molecular clouds ($n\gtrsim 10^3$--$10^7\,\text{cm}^{-3}$), the magnetic field increases with the density of the cloud \citep{Crutcher2012}. This is usually stated in form of a power law, $|\boldsymbol{B}|\propto \rho^{\kappa}$. If the cloud undergoes homologous compression, then magnetic flux ($\Phi=\pi R^2|\boldsymbol{B}|$) conservation implies $|\boldsymbol{B}|\propto R^{-2}$, while mass conservation gives $R\propto\rho^{1/3}$; therefore $|\boldsymbol{B}|\propto\rho^{2/3}$. On the other hand, if the magnetic field is strong, the structure of the cloud will be changed by the magnetic field. \citet{Fiedler} numerically showed that for ambipolar diffusion driven contraction $\kappa\approx 0.47$, which has also been seen observationally \citep{Crutcher1999}. However, \citet{Basu2000} showed that a better correlation was obtained by fitting $B \propto \sigma_v\sqrt{\rho}$.

The fifth row of Fig.~\ref{Figure 2} presents the dependence of plasma-$\beta$ with $a$. We see that for the MHD, medium-contraction model, the value of $\beta$ (the plasma $\beta$ is defined as the ratio between thermal pressure and magnetic pressure) is about 0.1 in the molecular regime \citep[for real molecular clouds the value of $\beta$ typically lies between 0.1--0.3; see][]{Crutcher2012,2016Federrath,Krumholz2019}, which implies a significant effect of the magnetic field ($\beta <1$) \citep{Plank2016}. Thus, we expect the value of $\kappa$ for our numerical experiment is 0.47 rather than $2/3$. We explore more about $B-n$ scaling relation and dependence on $\beta$ in Sec.~\ref{Sec 3.2}.

\subsection{Scaling relations}
\label{Sec 3.2}
\begin{figure}
    \centering
    \includegraphics[width=0.99\linewidth]{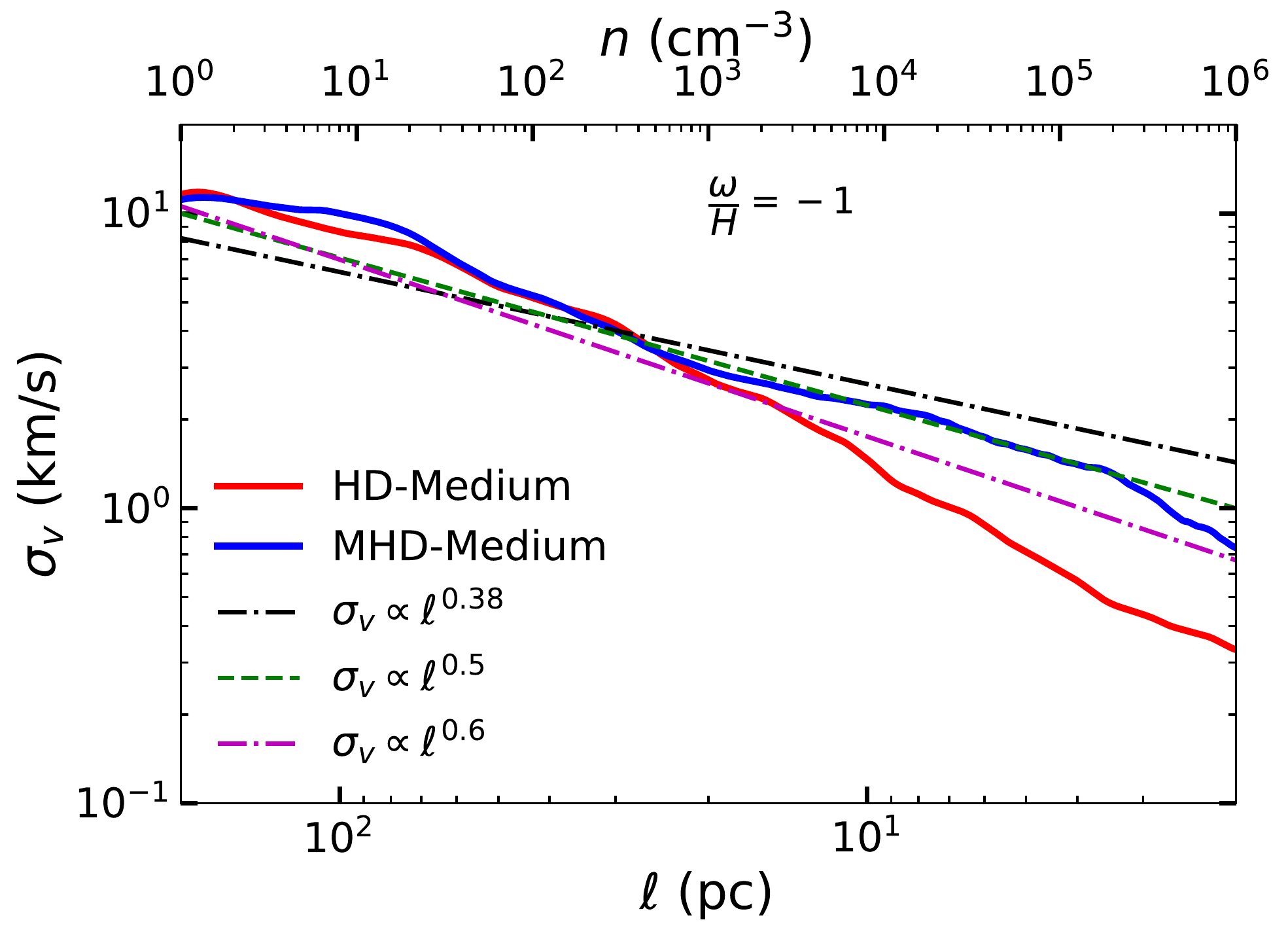}
    \caption{$\ell-\sigma_v$ scaling relation for the simulations with a medium-compression rate ($\omega/H=-1$). The red and blue solid lines correspond to HD and MHD simulations, respectively. The dashed lines are various observational scaling relations. The black dashed line is the $\sigma_v\propto\ell^{0.38}$ \citet{Larson1981} relation. The green and magenta dashed lines are $\sigma_v\propto \ell^{0.5}$ \citep{S87} and $\sigma_v\propto\ell^{0.5}$ \citep{B08} linewidth-Size scaling relations, respectively.}
    \label{Figure 3}
\end{figure}

From various surveys mentioned in Sec.~\ref{Sec 3.1.1} it is empirically established that the velocity dispersion ($\sigma_v$) in molecular clouds is correlated with the size of the cloud $\ell$ \citep{Larson1981, S87, Ossenkopf2002, Heyer2004, Roman-Duval2011}. Early observations predict that $\sigma_v$ is related with $\ell$ in a power law fashion $\sigma_v \propto \ell^{\xi}$. \citet{Larson1981} first calculated the value of $\xi$ to be $0.38$ (Larson relation). However, from various surveys, the most accepted linwidth-Size scaling relation in recent days is given by,
\begin{equation}
    \sigma_v \propto \ell^{0.5}, \label{velocity dispersion scaling}
\end{equation}
where $\ell$ is the size of the cloud in the unit of pc and the unit of $\sigma_v$ is km/s. This correlation also is reproduced by various numerical studies \citep{Kritsuk2007,Schmidt,Federrathetal2010,Federrath2013}. In Fig.~\ref{Figure 2} we added the scaling relation (equation~\ref{velocity dispersion scaling}) on top of the $\sigma_v-a$ lines. Only the simulation with medium compression provides a good match to the observed relation. Fig.~\ref{Figure 3} shows the scaling relation calculated from the medium compression models along with the \citep{Larson1981} relation (exponent $\xi=0.38$), Eq.~\ref{velocity dispersion scaling} and the B08 relation \citep{B08}, which has an exponent $\xi=0.5$. Result shows that the $\ell - \sigma_v$ relation from the medium compression model is consistent with the observational predictions. Moreover, Eq.~\ref{velocity dispersion scaling} fit very well to the MHD model in the molecular regime. This scaling relation is also consistent with the results from molecular cloud simulation driven by supernovae including self-gravity \citep{Padoan2017}.

\begin{figure}
    \centering
    \includegraphics[width=0.99\linewidth]{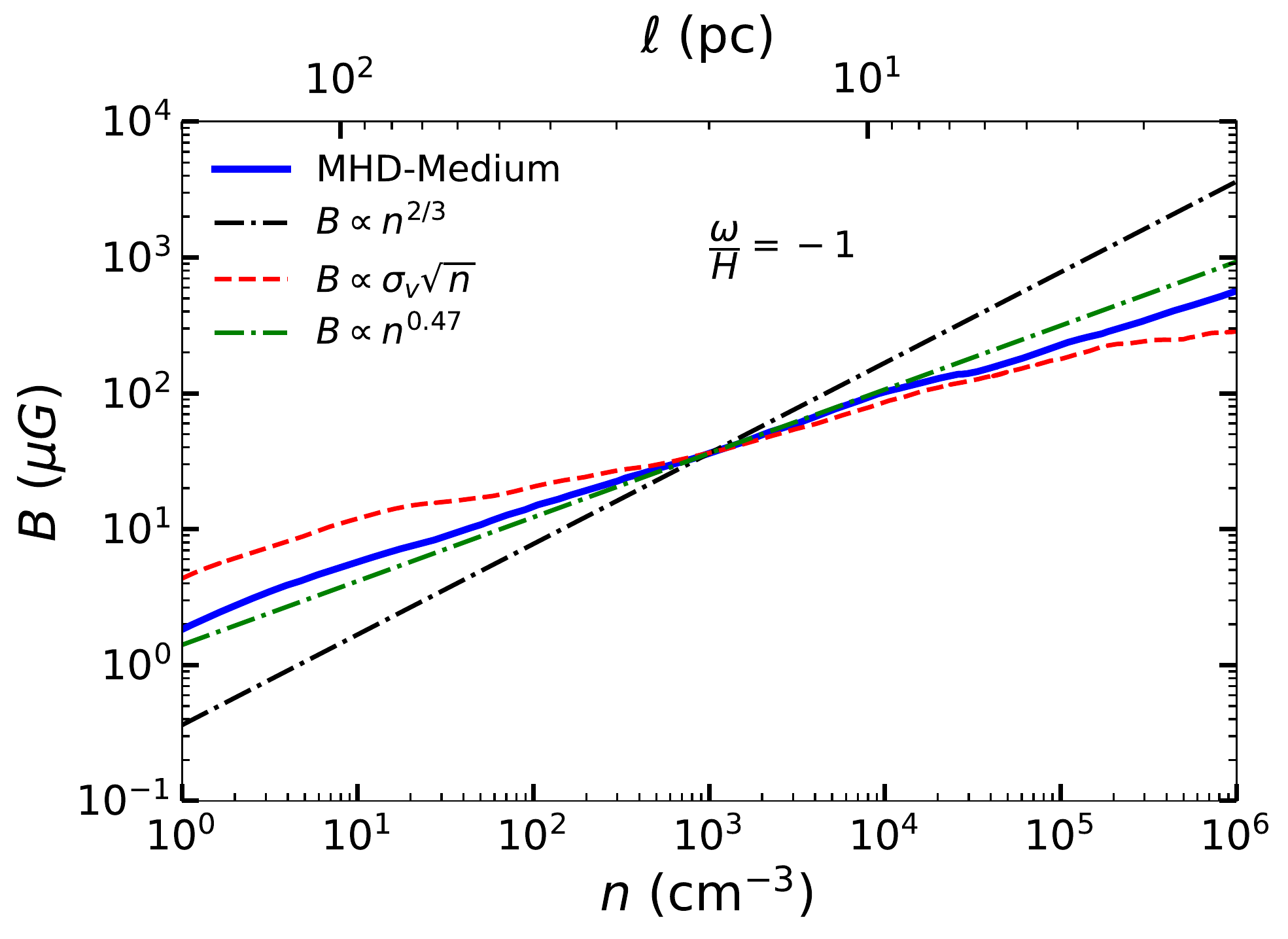}
    \caption{Relation between the magnetic field ($B$) and the density ($n$) for the medium-compression simulations ($\omega/H=-1$). The solid blue line corresponds to MHD simulation. The black dashed line presents the scaling relation $|\boldsymbol{B}|\propto n^{2/3}$ \citep{Crutcher2012}, resulting from homologous collapse of a cloud where the magnetic field is dynamically weak. The green dashed line represents the scaling relation $|\boldsymbol{B}|\propto n^{0.47}$, established by an ambipolar diffusion driven contraction model \citep{Fiedler}. Finally, the red dashed line shows the scaling relation $B \propto \sigma_v\sqrt{\rho}$ \citep{Basu2000}}
    \label{Figure 4}
\end{figure}

In Fig.~\ref{Figure 4}, we have plotted the dependence of $B$ with number density ($n$). We adopt the relation in the form of $|\boldsymbol{B}|=B_0\,(n/n_0)^{\kappa}$ and $|\boldsymbol{B}|=B_0(\sigma_v/\sigma_{v,0})(n/n_0)^{0.5}$, where $\kappa=2/3$ \citep{Crutcher2012} and 0.47 \citep{Fiedler} for two different models. Here $n_0=10^3\,\text{cm}^{-3}$, and $B_0$ and $\sigma_{v,0}$ are the magnetic field and velocity dispersion when the mean number density $n_0$. As pointed out earlier, the density-magnetic field correlation is valid in the high density regime $n \gtrsim 10^3\,\text{cm}^{-3}$, we choose the starting point of the plots at $n=10^3\,\text{cm}^{-3}$ and calculate the value of $B$ when $n=10^3\,\text{cm}^{-3}$ to find the constant $B_0$. We show the various scaling relations discussed in Sec.~\ref{Sec 3.1.4}. Since all of these scaling relations are valid only in the high-density regime ($n\sim10^3$--$10^6\,\text{cm}^{-3}$), we expect our simulation results to be consistent with these relations only in the higher-density regime. Fig.~\ref{Figure 4} shows good agreement of the MHD medium-compression model with the theoretical and observational estimates of the $B-n$ relation. As pointed out in Sec.~\ref{Sec 3.1.4}, $\beta<1$ implies a significant magnetic influence on the evolution of the cloud, thus $\kappa$ value will be close to 0.47. Moving-mesh simulations of isothermal, self-gravitating molecular cloud cores \citep{Mocz2017} find $\kappa \approx 1/2$, when the magnetic field is strong ($\beta < 1$), consistent with the simulations presented here. The MHD-Medium compression result approximately follows the $n^{0.47}$ curve. It also fits well to the $\sigma_v\sqrt{n}$ curve in the molecular regime. All of the above results suggest that the $\omega/H=-1$ contraction model produces reasonable cloud parameters as a natural outcome of molecular cloud formation by compression out of the warm atomic phase.

\subsection{Morphological features}
\label{Sec 3.3}
\begin{figure*}
  \includegraphics[width=1.0\linewidth]{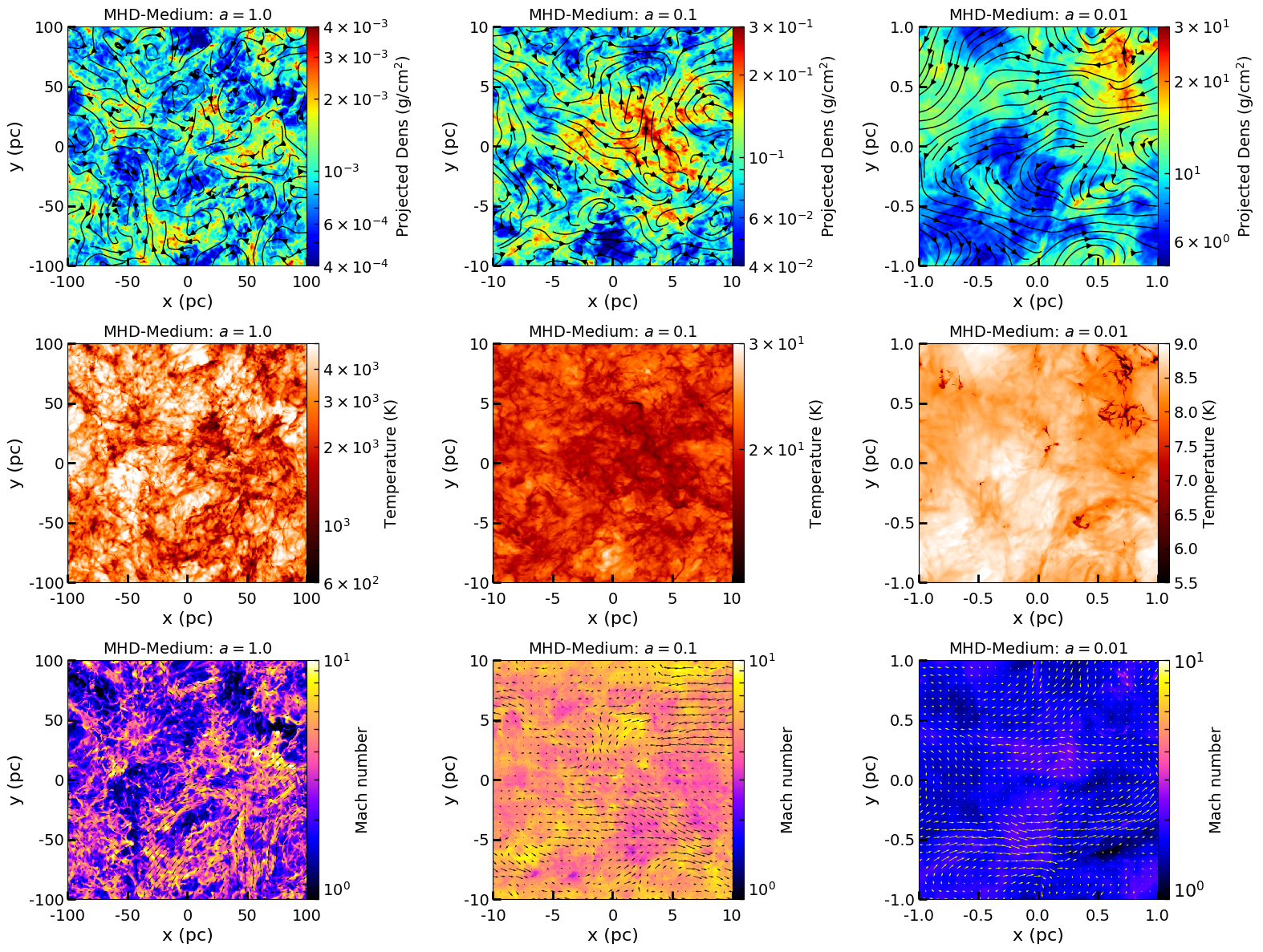}
  \caption{The first row shows the projected density in the $x$--$y$ plane for the MHD-Medium simulation. The second and third-row represent the line-of-sight, density-weighted mean temperature and Mach number, respectively. The first, second and third column correspond to $a=1.0$, $0.1$, and $0.01$. The streamlines in the density projections represent the projected magnetic field in the $x$--$y$ plane. The velocity field is shown as arrows in the Mach number projections (bottom panels).}
  \label{Figure 5}
\end{figure*}

Fig.~\ref{Figure 5} displays a spatial representation of density (top panel), temperature (middle panel) and Mach number (bottom panel) at $a=1.0$ (left), $a=0.1$ (middle) and $a=0.01$ (right) for the MHD-Medium model. We have plotted the local magnetic field lines projected onto the $x$--$y$ plane on top of the density projections, and local velocity field vectors in the Mach number projections. In the density projections, we see that initially ($a=1$), the large-scale turbulence driving sets the density contrasts ranging over one order of magnitude and the local magnetic fields are quite random. However, at a later times ($a=0.1$ and $0.01$), when the strength of the magnetic fields increase, the density contrasts decrease (ranging over a factor of 3). This is because magnetic fields reduce the density contrasts, due to additional magnetic pressure parameterized by the plasma-$\beta$ \citep{Molina, Federrath2012}. The magnetic field directions at this epoch are more regular, which is consistent with the strong-field predictions, i.e., for a strong field, the field lines should be smoother \citep{Crutcher2012}. The temperature (second row in Fig.~\ref{Figure 5}) also shows similar behavior. Initially ($a=1$), the temperature fluctuations cover almost four orders of magnitude, while at $a=0.01$, the temperature fluctuations are very small and gas is nearly isothermal at $T\sim5$--$10\,\mathrm{K}$.

One interesting point to notice is the correlation between density and temperatures. In Fig.~\ref{Figure 5}, we see that the correlations are very prominent. Higher densities have lower temperature as expected due to cooling. The correlations between density and Mach number are quite weak. However, Fig.~\ref{Figure 5} shows that high-density regions exhibit a lower Mach number on average, as a result of the velocity dispersion -- size relation discussed above.

\subsection{Density dispersion -- Mach number relation}
\label{Sec 3.4}
\begin{figure*}
\centerline{
\def\arraystretch{1.0}
\setlength{\tabcolsep}{0.0pt}
\begin{tabular}{lcr}
  \includegraphics[width=0.5\linewidth]{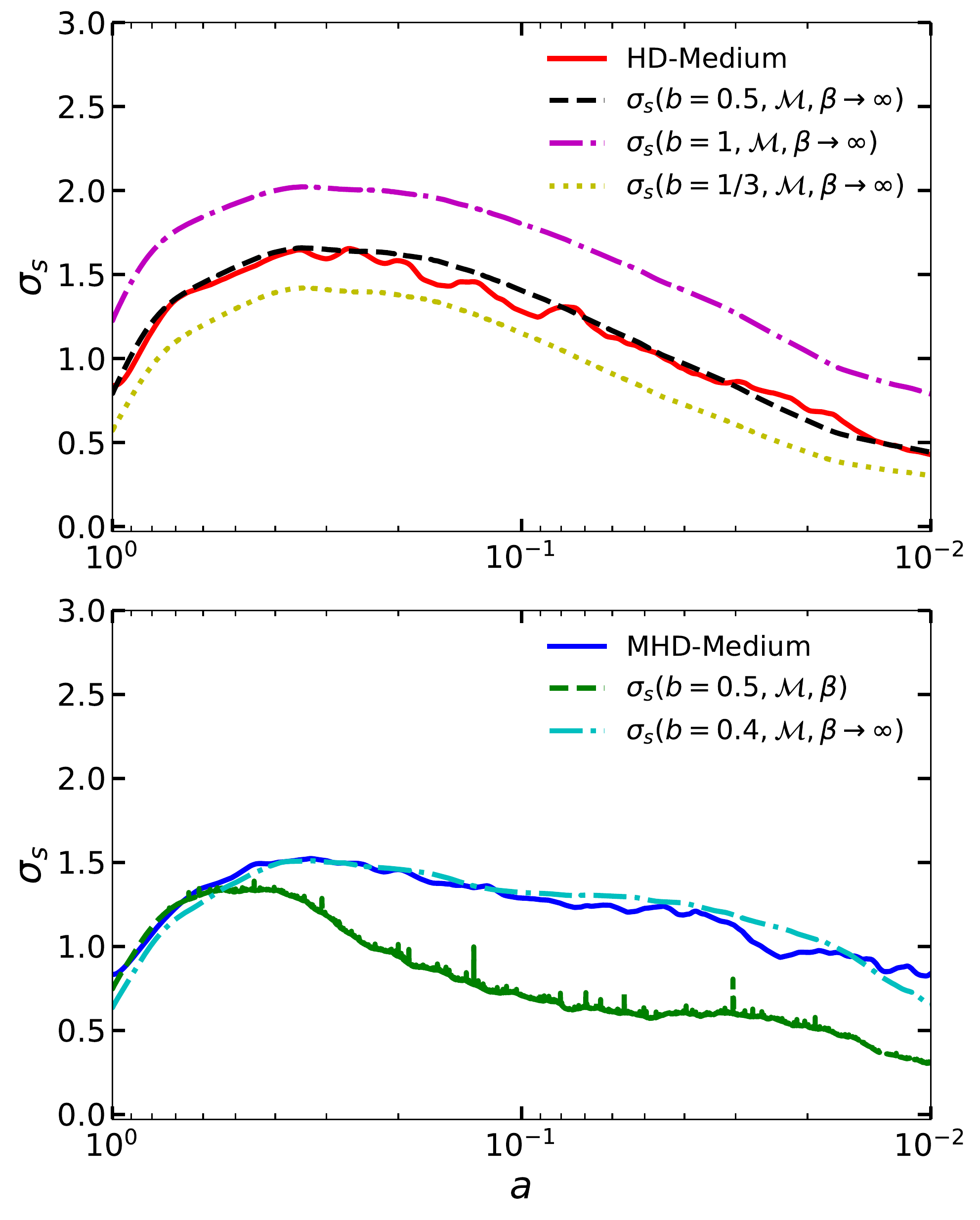} &
  \includegraphics[width=0.5\linewidth]{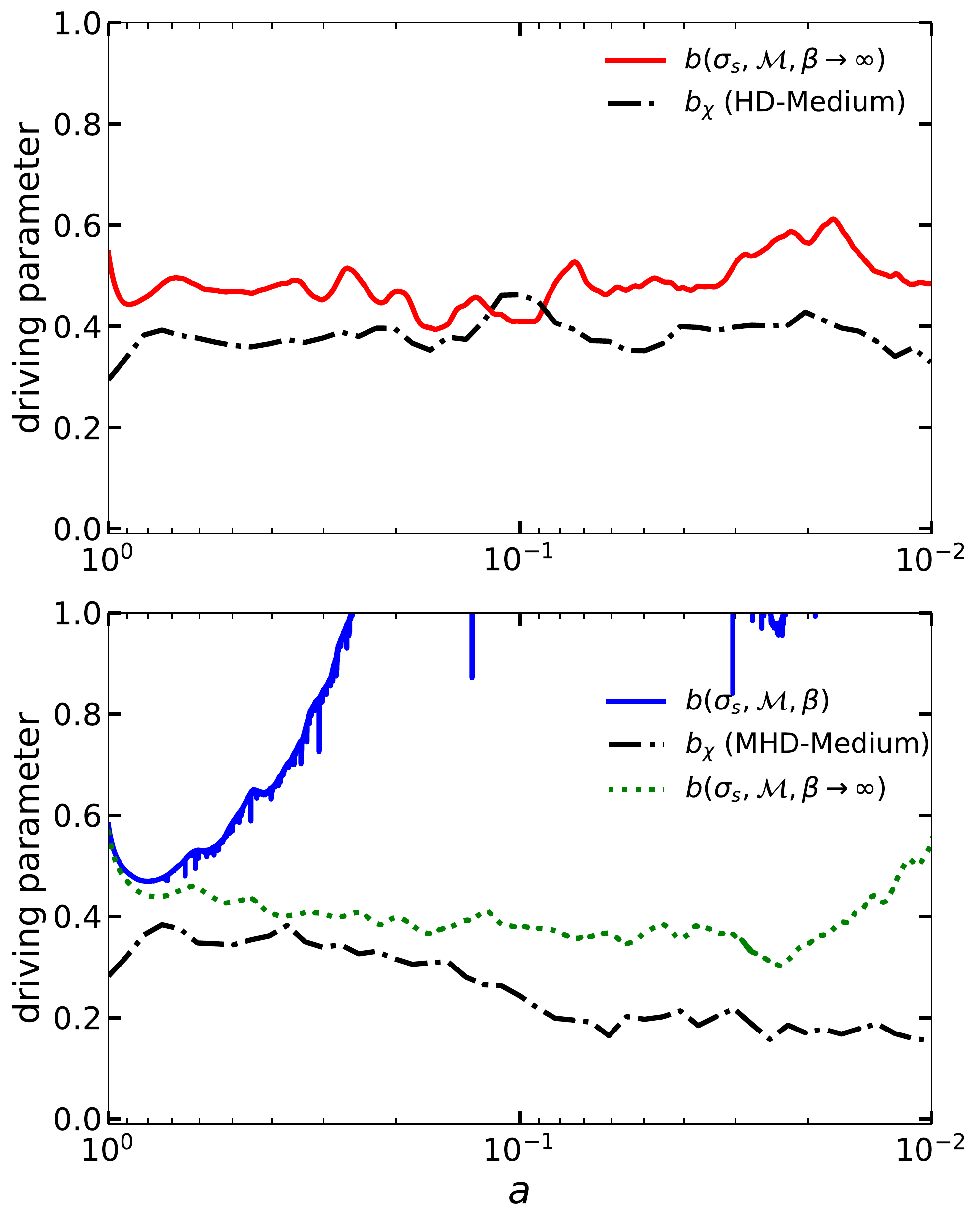}
\end{tabular}}
  \caption{Evolution of standard deviation of log-density ($\sigma_s$) (left-hand panels) and driving parameter $b$ and $b_\chi$ (right-hand panels) as a function of the scale factor $a$ for HD-Medium (top panels) and MHD-Medium (bottom panels). In the upper-left panel, the red solid line represents the evolution of $\sigma_s$ for HD simulation and the black dashed line is the corresponding theoretical prediction, Eq.~(\ref{log density variance HD}) with $b=0.5$. The magenta and yellow lines in the upper-left panel correspond to the theoretical predictions for purely solenoidal ($b\sim 1/3$ and purely compressive ($b\sim 1$) driving. The bottom-left panel presents the result for MHD-Medium (blue solid line) and the green line corresponds to the theoretical model, Eq.~(\ref{log density variance MHD}) with $b=0.5$. The cyan line in the bottom panel is the HD limit ($\beta \rightarrow \infty$) of the MHD model (Eq.~\ref{log density variance MHD}) with $b=0.4$, which fits the data well. In the upper-right panel, the red solid line corresponds to the evolution of the driving parameter ($b$) that has been calculated using Eq.~\eqref{b} with $\beta \rightarrow \infty$ and the black dashed-dotted line is $b_\chi$ calculated using Eq.~\eqref{driving parameter}. In the bottom-right panel, the blue line is $b$ calculated from Eq.~\eqref{b} and the black dashed dotted line is $b_\chi$. The green dotted line corresponds to Eq.~\eqref{b} with $\beta \rightarrow\infty$}.
  \label{Figure 6}
\end{figure*}
In supersonic, isothermal turbulence, the density fluctuations approximately follow a log-normal distribution \citep{Vazquez1994}, assuming that the local density fluctuations and velocities are uncorrelated \citep{Passot1998,Federrathetal2010,Federrath2015,Kritsuk2017}. However, if the gas is not isothermal, we do not expect a log-normal distribution for density fluctuations, as the local density and Mach number are correlated \citep{Passot1998,Gazol2013,Kortgen2019}. Various theoretical and numerical studies \citep{Padoan1997,Passot1998,Kowal2007,Federrath2008} have established that the density dispersion and Mach number follow the relation,
\begin{equation}
    \sigma_s(b,\mathcal{M})=\left[\ln(1+b^2\mathcal{M}^2)\right]^{1/2}, \label{log density variance HD}
\end{equation}
where $\sigma_s$ is the variance of the logarithmic density contrast, $s=\ln(\rho/\langle \rho \rangle)$. \citet{Federrath2008,Federrathetal2010} pointed out that the parameter $b$ is a function of how the turbulence is driven and found $b\approx1/3$ for purely solenoidal (divergence-free) driving and $b\approx1$ for purely compressive (curl-free) driving. For magnetized turbulence the \mbox{$\sigma_s$--$\mathcal{M}$} relation is modified by magnetic pressure and was analytically derived by \citet{Padoan2011} and \citet{Molina}. For $B\propto\rho^{1/2}$ they find,
\begin{equation}
    \sigma_s(b,\mathcal{M}, \beta)=\left[\ln\left(1+b^2\mathcal{M}^2\frac{\beta}{\beta +1}\right)\right]^{1/2}, \label{log density variance MHD}
\end{equation}
where $\beta$ is the ratio of thermal to magnetic pressure (denoted plasma-$\beta$) of the magnetized flow. Equation~\eqref{log density variance MHD} is a more general form, as we can see in the limit of $\beta \rightarrow \infty$ (hydrodynamic limit), Equation~\eqref{log density variance MHD} and \eqref{log density variance HD} are identical.

For given $\sigma_s$, $\mathcal{M}$, and $\beta$, Eqs.~\eqref{log density variance MHD} can be used to calculate the effective driving parameter $b$ of the turbulence, i.e.,
\begin{equation}
    b=\left[\frac{\beta+1}{\beta \mathcal{M}^2}\left(\exp(\sigma_s^2)-1\right)\right]^{1/2}. \label{b}
\end{equation}
In the limit $\beta\rightarrow\infty$, Eq.~\eqref{b} gives the effective driving parameter for HD turbulence.

An alternative way of estimating $b$ is to consider the mixture of turbulent modes in the velocity field. For this purpose, \citet{PanEtAl2016} define the compressive ratio
\begin{equation}
    \chi=\vcomp/\vsol, \label{compressive ratio}
\end{equation}
where $\vcomp$ and $\vsol$ are the compressive and soleloidal components of the velocity field, respectively. An effective driving parameter ($b_{\chi}$) can then be defined as
\begin{equation}
    b_{\chi}=\sqrt{\frac{\chi}{1+\chi}}. \label{driving parameter}    
\end{equation}
The velocity components $\vcomp$ and $\vsol$ can be determined via a Helmholtz decomposition in Fourier space. Thus, from the velocity power spectra one can determine $b_{\chi}$ using Eqs.~\eqref{compressive ratio} and~\eqref{driving parameter}. However, $b$ in Eq.~\eqref{b} may not be the same as $b_{\chi}$ in Eq.~\eqref{driving parameter}, but \citet{PanEtAl2016} and \citet{JinEtAl2017} argue and show that they are usually fairly similar.

In Fig.~\ref{Figure 6} we show $\sigma_s$ (left panels) and the driving parameter ($b$, $b_{\chi}$; right panels) as a function of $a$. The top panels represent the evolution for the HD-Medium simulation and the bottom panels show the same for the MHD-Medium simulation. Since in the driving phase (before contraction), the turbulence is driven by a solenoidal acceleration field ($b\sim1/3$), and contraction acts primarily as a compressive acceleration, the resulting driving during the contraction phase is a mixture of solenoidal and compressive components ($1/3<b<1$) \citep[Fig.~8]{Federrathetal2010}. For our case, we find that \mbox{$b\approx 0.4$--$0.5$} provides a good fit in the case without magnetic fields (HD-Medium in the top panels of Fig.~\ref{Figure 6}). We find that the driving parameter $b$ computed using Eq.~\ref{b} with $\beta \rightarrow\infty$ and $b_{\chi}$ computed from the Helmholtz-decomposed velocity field are in good agreement, with $b_\chi\sim0.4$, somewhat smaller than $b\sim0.5$.

In the bottom-left panel of Fig.~\ref{Figure 6}, we show the evolution of $\sigma_s$ for the MHD-Medium simulation. The turquoise line represents the prediction from the theoretical model in the HD limit (Eq.~\ref{log density variance HD}) with $b=0.4$, while the green line is for the MHD model with the $\beta$-parameter included and $b=0.5$ (Eq.~\ref{log density variance MHD}), for comparison. The bottom-right panel shows the evolution of $b$ and $b_\chi$ for the MHD-Medium model. The blue solid line corresponds to $b$ directly computed from Eq.~\eqref{b} and the black dashed-dotted line is $b_{\chi}$ computed using Eq.~\eqref{driving parameter}. The green dotted line corresponds to the evolution of $b$ calculated using Eq.~\eqref{b} with $\beta\rightarrow\infty$.

We see that the MHD model (Eq.~\ref{driving parameter}) does not provide a good fit to the data and suggests $b>1$ for $a\lesssim0.2$. This is because Eq.~\ref{log density variance MHD} breaks down as the Alfv\'en Mach number drops below $\sim2$. Indeed, \citet{Molina} have shown that this \mbox{$\sigma_s$--$\mathcal{M}$} relation (Eq.~\ref{log density variance MHD}) does not work for $\mathcal{M}_{\mathrm{A}} \lesssim 2$, and in our case, $a\lesssim 0.2$ is when this transition to strongly sub-Alfv\'enic turbulence occurs (c.f., the bottom panel of Fig.~\ref{Figure 2}).

Despite these caveats, we find that using the purely hydrodynamical version of the \mbox{$\sigma_s$--$\mathcal{M}$} relation, as well as $b_\chi$, suggest a driving parameter of $\sim0.3$--$0.4$, with $b_\chi$ giving a somewhat lower effective driving parameter close to purely solenoidal or less (\mbox{$b_\chi\sim0.2$--$0.3$}) than $b\sim0.4$ from the HD-version of the \mbox{$\sigma_s$--$\mathcal{M}$} relation.

Overall, these results suggest that contracting MHD turbulence has an effectively solenoidal to naturally-mixed driving parameter (\mbox{$b\sim0.3$--$0.4$}), while contracting HD turbulence is naturally mixed to slightly compressive (\mbox{$b\sim0.4$--$0.5$}).

\subsection{The density PDFs for different compression models}\label{Sec 3.5}

\begin{figure*}
\centerline{
\def\arraystretch{1.0}
\setlength{\tabcolsep}{0.0pt}
\begin{tabular}{lcr}
  & \includegraphics[width=0.33\linewidth]{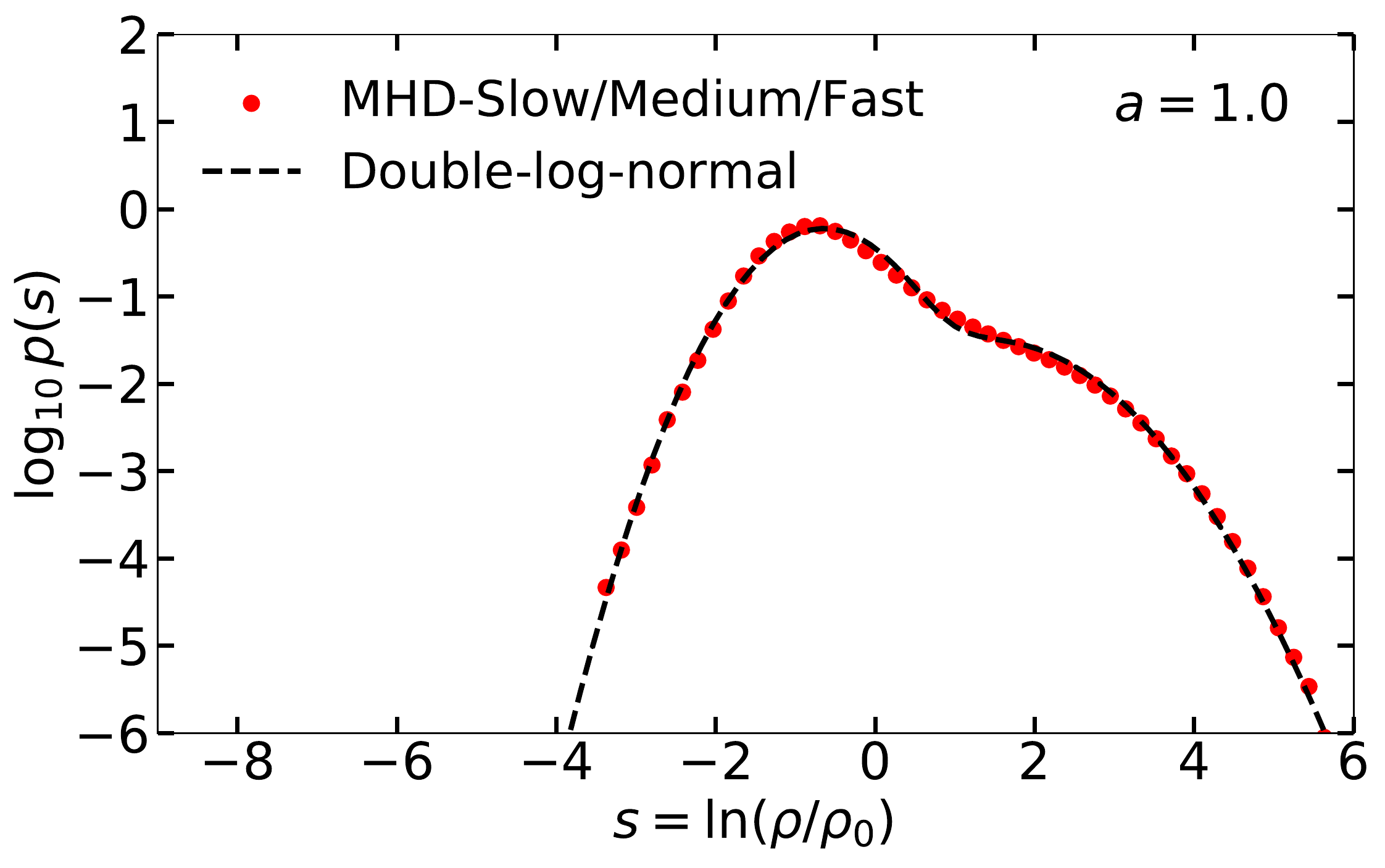} &\\
  \includegraphics[width=0.33\linewidth]{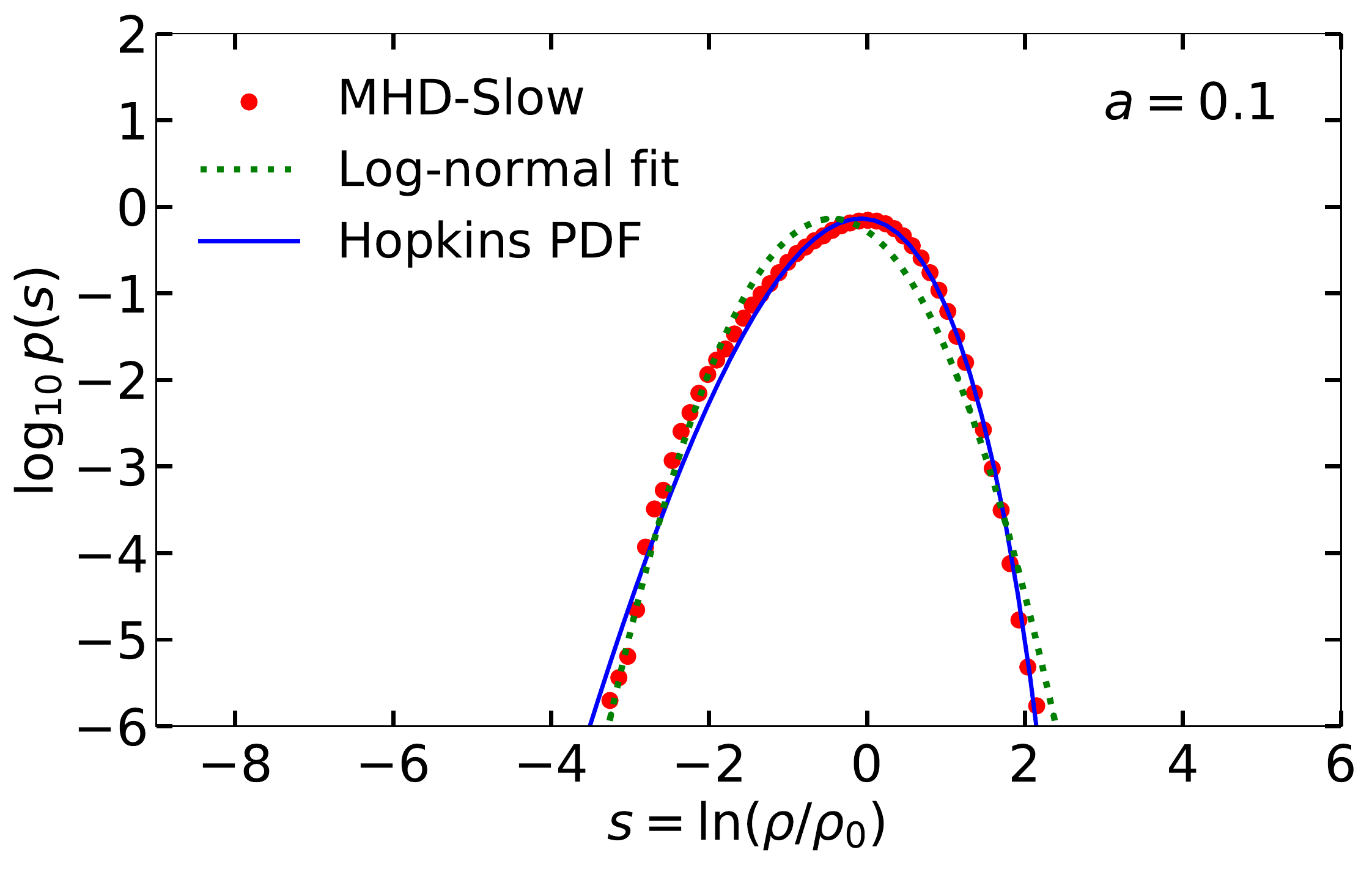} &
  \includegraphics[width=0.33\linewidth]{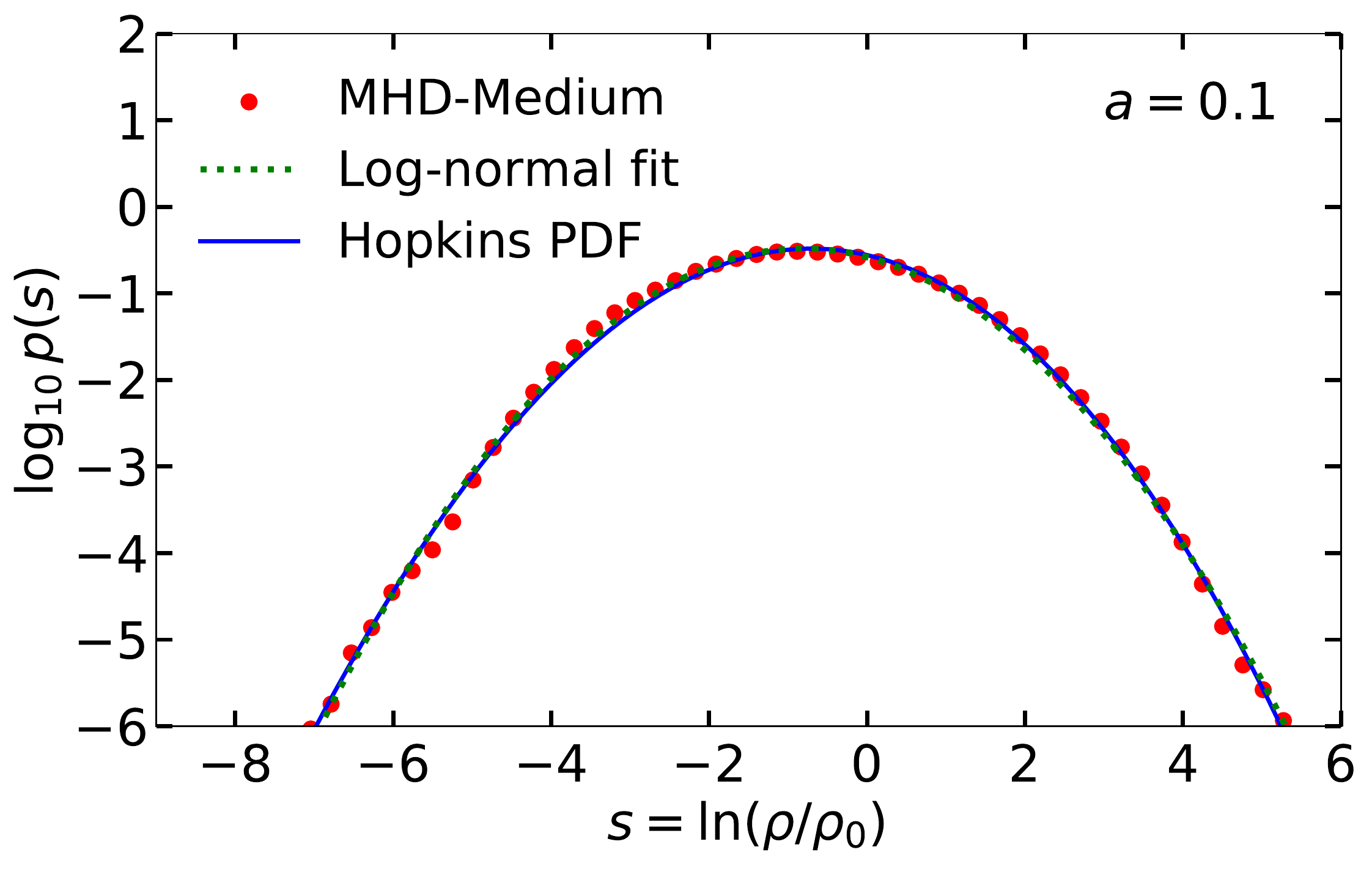} &
  \includegraphics[width=0.33\linewidth]{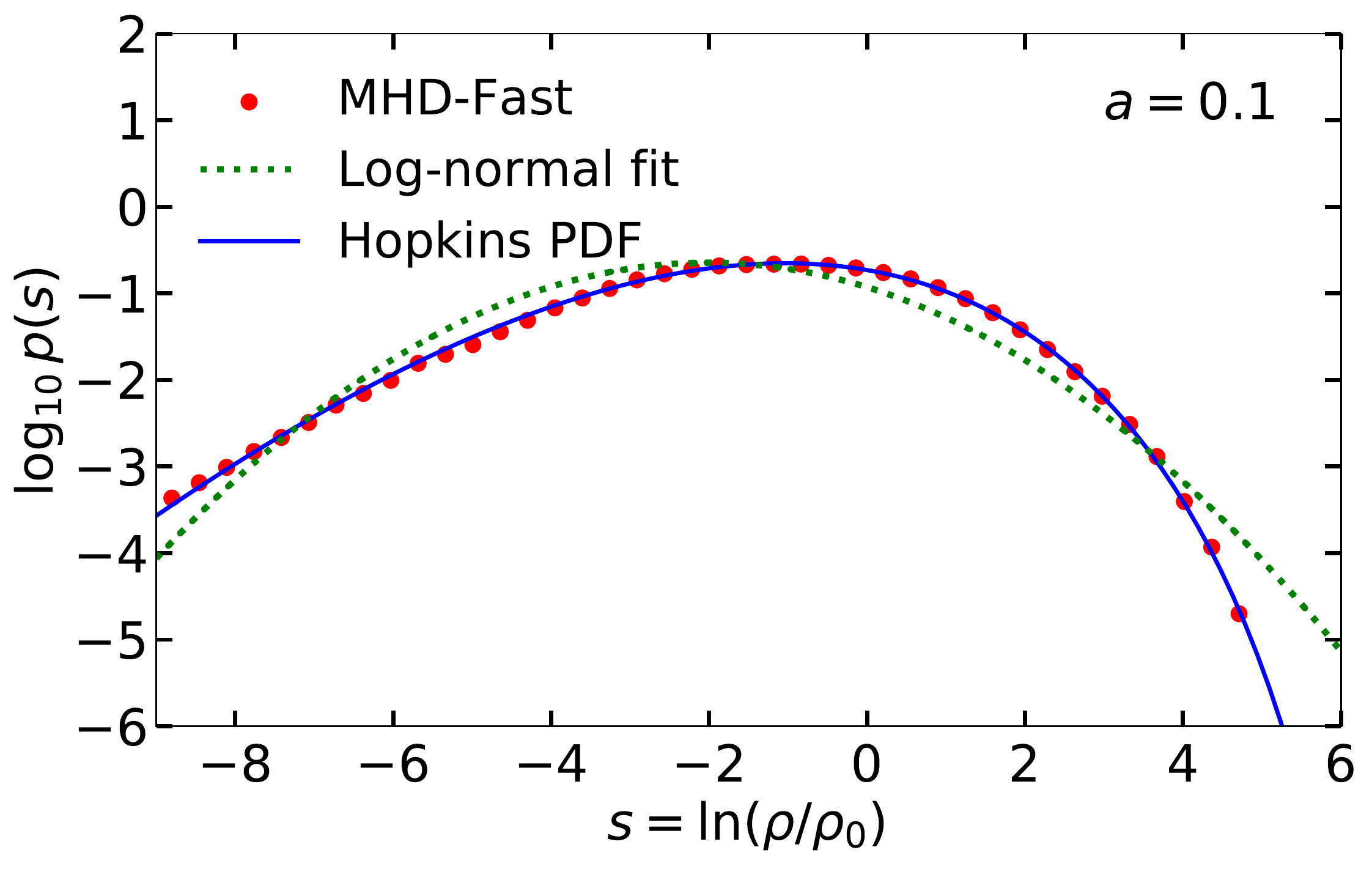} \\
  \includegraphics[width=0.33\linewidth]{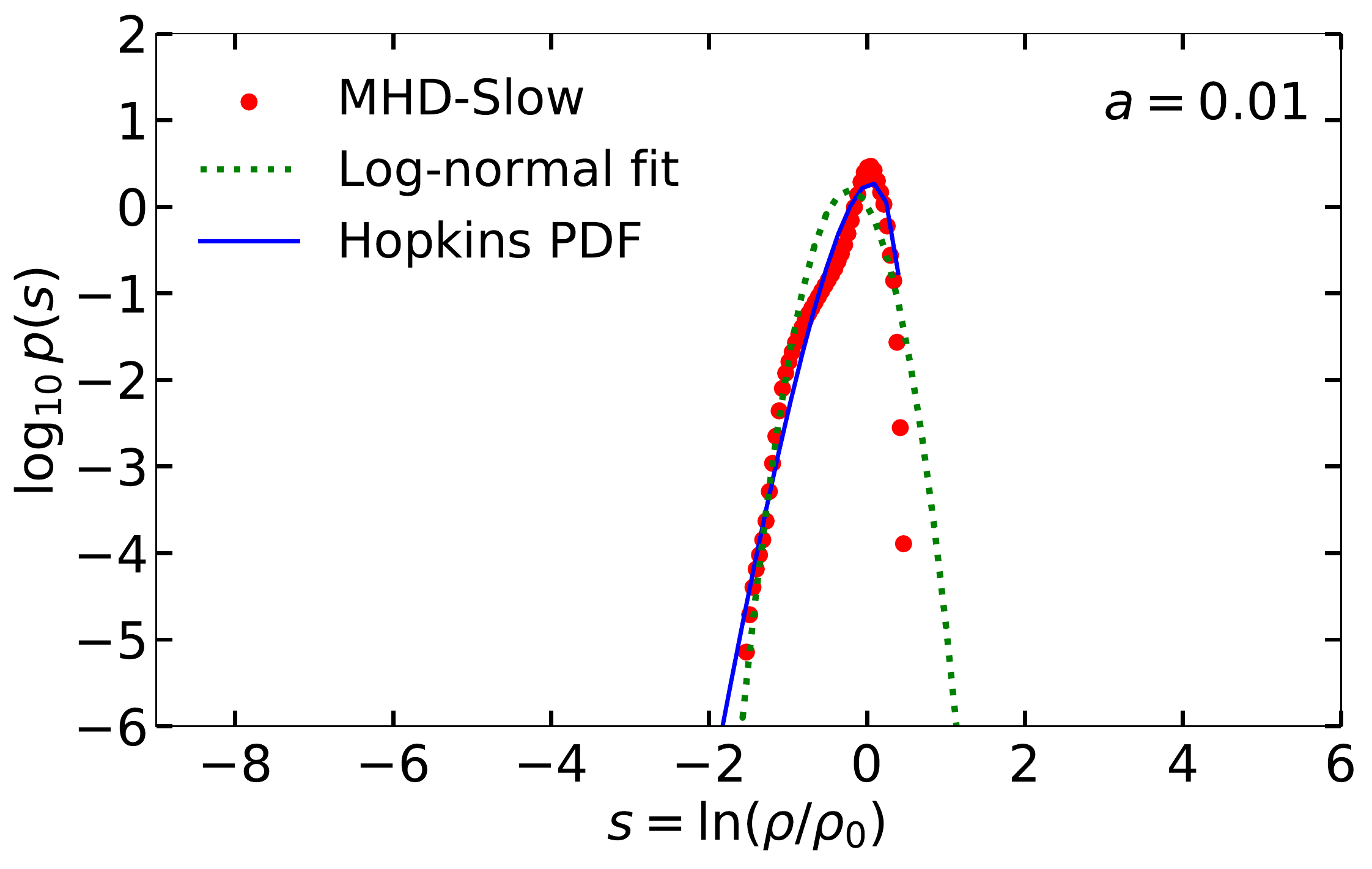} &
  \includegraphics[width=0.33\linewidth]{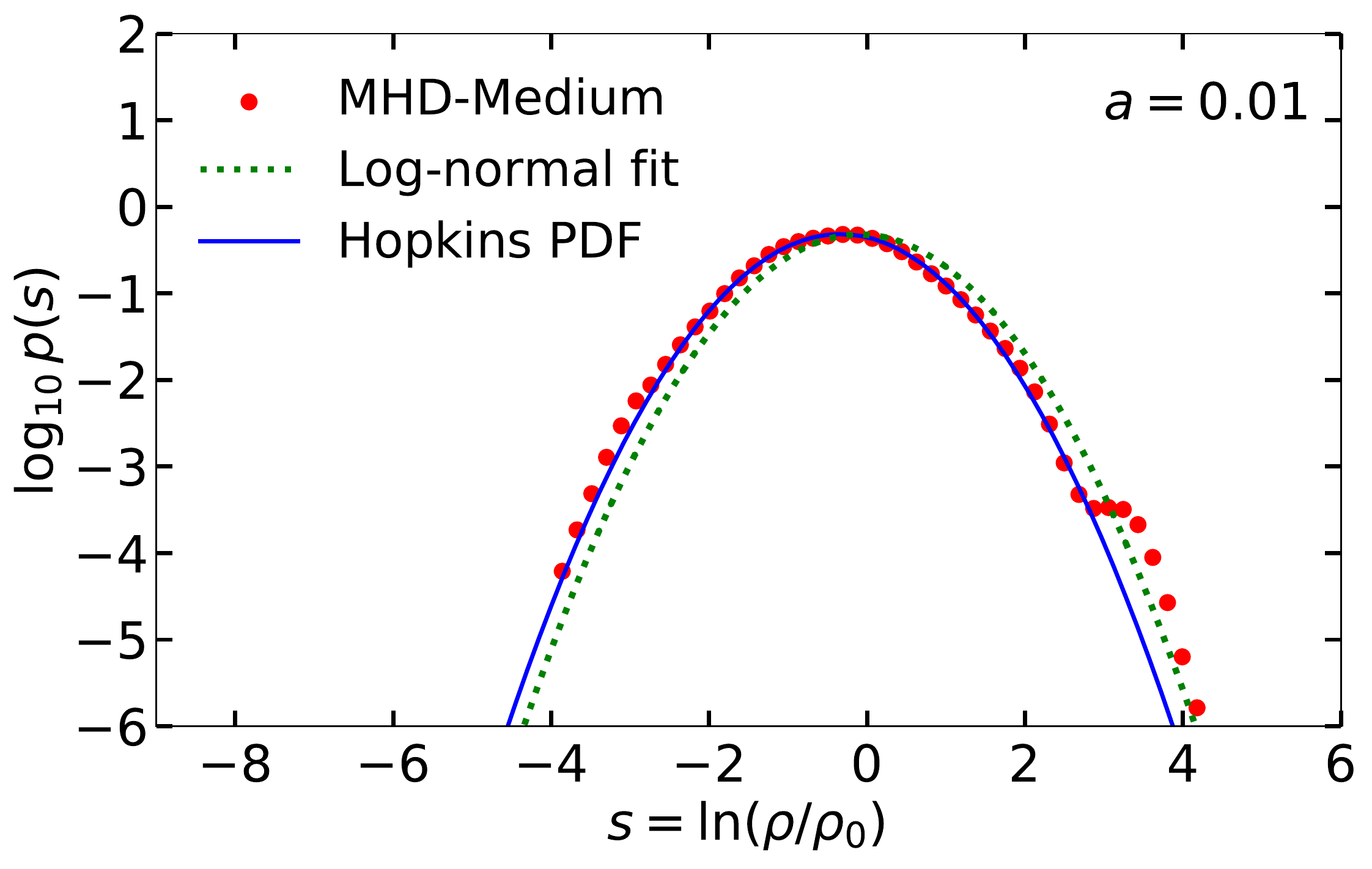} &
  \includegraphics[width=0.33\linewidth]{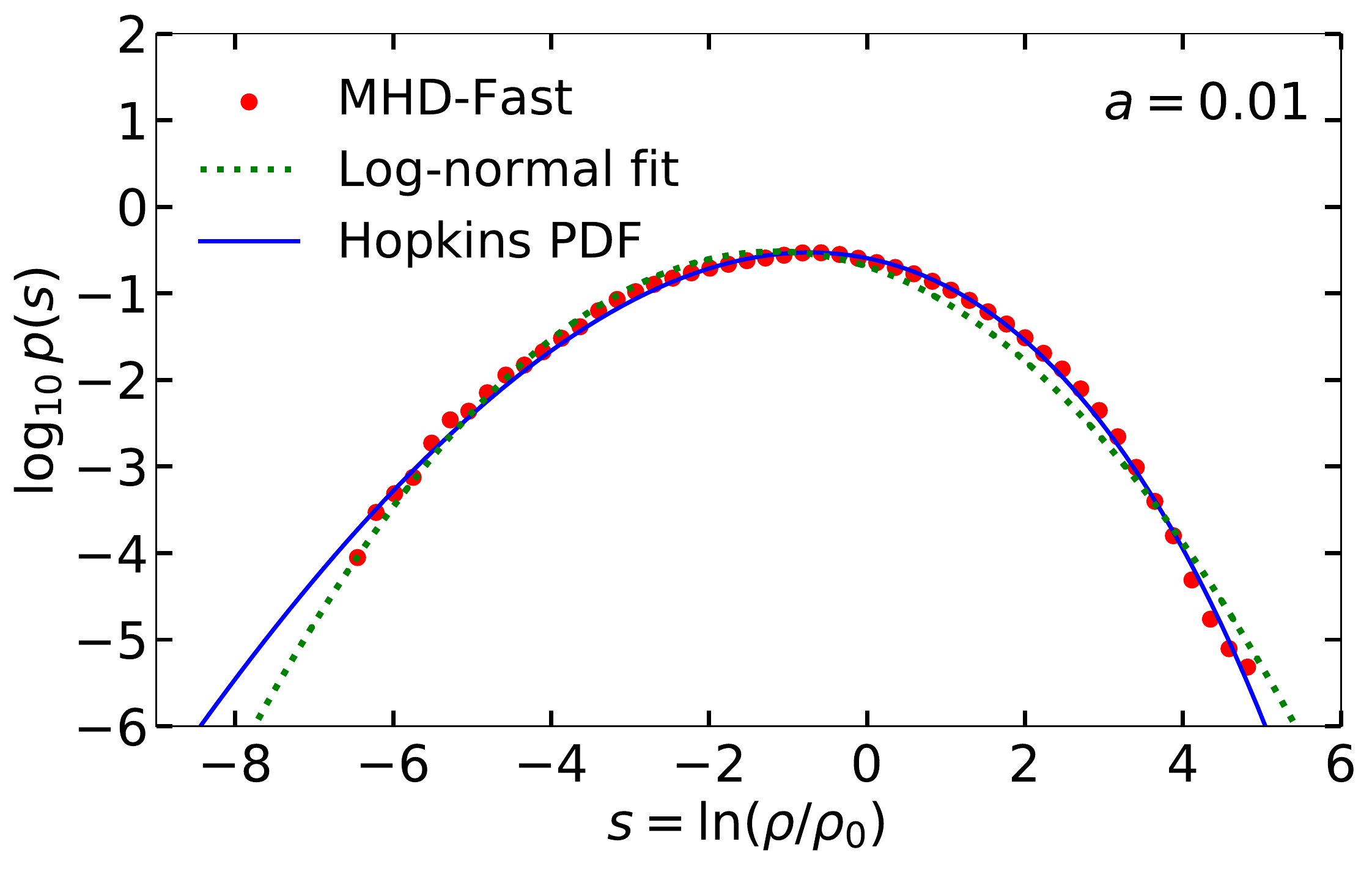}
\end{tabular}}
  \caption{PDFs of logarithmic density contrast $s\equiv \mathrm{ln}(\rho/\rho_0)$ for different contraction models. In the first row, we show the density PDF at $a=1.0$ (just after the turbulence is fully developed and just before the contraction starts). The second and third row correspond to the density PDFs at $a=0.1$ and $0.01$, for different compression models, respectively. Different columns (from left to right) correspond to MHD-Slow, MHD-Medium, and MHD-Fast. In each panel, the red points are the simulation PDFs. The black dashed line in the first panel is the double-log-normal fit (Eq.~\ref{double-log-normal}) to the numerical PDF at $a=1.0$. In all subsequent panels, the green dotted lines represent log-normal fits (Eq.~\ref{log-normal PDF}) and the blue solid lines show fits with the intermittency PDF model (Eq.~\ref{Hopkins pdf}).}
  \label{Figure 7}
\end{figure*}

\begin{table*}
    \caption{Fitting parameters for the double-log-normal function at $a=1$}
    \label{Table 2}
    \begin{tabular*}{\textwidth}{c @{\extracolsep{\fill}} ccccccc}
    \hline
    Model & $\omega/H$ & a & $N$ & $\sigma_{s1}$ & $\sigma_{s2}$ & $s_{01}$ & $s_{02}$ \\
    \hline
    MHD-Slow/Medium/Fast & -1.0 & 1.0 & $1.670\pm 0.017$ & $0.618\pm 0.006$ & $0.927\pm 0.016$ & $-0.644\pm0.021$ & $1.423\pm 0.079$ \\
    \hline
    \end{tabular*}
\end{table*}
\begin{table*}
    \caption{Statistical measures of the fitting parameters at $a=0.1$ and $0.01$}
    \label{Table 3}
    \begin{tabular*}{\textwidth}{c @{\extracolsep{\fill}} ccccccc}
    \hline
    \hline
    Model & $\omega/H$ & a & Sim $\sigma_{s,V}$ & log-normal $\sigma_{s,V}$ & log-normal $s_0$ & Hopkins $\sigma_{s,V}$ & Hopkins $\theta$  \\
    \hline
    MHD-Slow & $-0.1$ & $0.1$ & $0.833$ & $0.831\pm 0.010$ & $-0.092\pm0.040$ & $0.823\pm 0.012$ & $0.028\pm 0.011$ \\
    MHD-Slow & $-0.1$ & $0.01$ & $0.191$ & $0.254\pm 0.009$ & $-0.226\pm0.034$ & $0.221\pm 0.004$ & $0.050\pm 0.002$ \\
    \hline
    MHD-Medium & $-1.0$ & $0.1$ & $1.278$ & $1.212\pm 0.003$ & $-0.812\pm0.013$ & $1.212\pm 0.003$ & $0.012\pm 0.002$ \\
    MHD-Medium & $-1.0$ & $0.01$ & $0.833$ & $0.831\pm 0.010$ & $-0.092\pm0.040$ & $0.823\pm 0.012$ & $0.028\pm 0.011$ \\
    \hline
    MHD-Fast & $-10.0$ & $0.1$ & $1.778$ & $1.762\pm 0.029$ & $-2.010\pm0.079$ & $1.807\pm 0.004$ & $0.207\pm 0.002$ \\
    MHD-Fast & $-10.0$ & $0.01$ & $1.392$ & $1.312\pm 0.009$ & $-1.165\pm0.031$ & $1.346\pm 0.006$ & $0.077\pm 0.003$ \\
    \hline
    \end{tabular*}
\end{table*}

As pointed out in Sec.~\ref{Sec 3.4}, for an idealized isothermal turbulence, the density contrast $(\rho/\rho_0)$ follows an approximately log-normal distribution, a Gaussian distribution in the logarithmic density contrast $s\equiv\ln(\rho/\rho_0)$,
\begin{equation}
    p_\mathrm{LN}(s)=\frac{1}{\sqrt{2\pi \sigma_{s}^2}}\exp\left(-\frac{(s-s_0)^2}{2\sigma_{s}^2}\right). \label{log-normal PDF}
\end{equation}
The log-normal PDF contains two parameters: 1) the variance of logarithmic density $\sigma_s$ and 2) the mean $s_0$, which is related to the variance by $s_0=\sigma_s^2/2$ because of mass conservation \citep{PadoanEtAl1997,Federrath2008,Hopkins2013}. Although the log-normal model of the density PDF for isothermal turbulence is well established, this is not obvious for non-isothermal turbulence. Previous studies show that the PDF tends to depart significantly from the log-normal form given by Eq.~\ref{log-normal PDF}, if the gas is non-isothermal \citep{Vazquez1994,Passot1998,Wada2001,Kritsuk2002,Li2003,Audit2005,Hennebelle_Audit2007,Seifried2011,Molina,Hopkins2013,Gazol2013,Federrath_Banerjee2015}. A model for capturing the non-lognormality of density fluctuation is the intermittency model presented by \citet{Hopkins2013}, physically motivated by the quantized log-Poisson statistics for velocity structure functions ($S_p(\boldsymbol{R})\equiv \langle\delta v(\boldsymbol{R}^p\rangle$) \citep{Dubrulle1994,She1995}. The form of the density PDF in case is \citep{Hopkins2013,Federrath_Banerjee2015},
\begin{multline}
    p_\mathrm{HK}(s)=I_1\big(2\sqrt{\lambda w(s)}\big)\exp[-(\lambda+w(s))]\sqrt{\frac{\lambda}{\theta^2 w(s)}}, \\
    \lambda\equiv \frac{\sigma_s^2}{2\theta^2}, \,\,\,\, w(s)\equiv \frac{\lambda}{1+\theta}-\frac{s}{\theta} \,\,\, (w\geq 0), \label{Hopkins pdf}
\end{multline}
where $I_1(x)$ is the 1st-order \textit{modified Bessel function} of the first kind. Eq.~(\ref{Hopkins pdf}) contains two parameters: 1) the standard deviation $\sigma_s$ and 2) the intermittency parameter $\theta$. In the zero-intermittency limit ($\theta \rightarrow 0$), Eq.~(\ref{Hopkins pdf}) simplifies to the log-normal PDF, Eq.~(\ref{log-normal PDF}).

\citet{Hopkins2013}, \citet{Federrath2013}, and \citet{Federrath_Banerjee2015} showed that this intermittency model for the density PDF fits well to the density PDFs of idealized non-isothermal turbulence simulations with a large range of Mach numbers, magnetic field strengths, and turbulence driving parameters. Thus, here we are motivated to examine whether this also true for our non-isothermal contraction model with radiative heating and cooling. 

Fig.~\ref{Figure 7} shows the density PDFs of $s\equiv\ln (\rho/\rho_0)$ obtained in our simulations with different contraction rate. In the first row of Fig.~\ref{Figure 7}, we show the density distribution at $a=1$ (i.e., when the turbulence is fully developed and immediately before the contraction starts). The second and third row of Fig.~\ref{Figure 7} correspond to the density distribution at $a=0.1$ and $0.01$ for different compression models, respectively. Different columns (from \textit{left} to \textit{right}) correspond to MHD-Slow, MHD-Medium, and MHD-Fast compression models. In each panel, the red points are the density PDFs from the simulation data.

At $a=1$ (top panel in Fig.~\ref{Figure 7}), we see that the density distribution is divided into two phases: 1) diffuse warm gas, and 2) dense cold gas. Thus, natural motivation indicates that each of these phases may follow a separate log-normal distribution. Thus, for the entire density range, we fit a double-log-normal PDF \citep[e.g.,][]{Dawson2015},
\begin{multline}
    p_\mathrm{DLN}(s)=\frac{N}{\sqrt{2\pi\sigma_{s1}^2}}\exp\left(-\frac{(s-s_{01})^2}{2\sigma_{s1}^2}\right) + \\ \frac{1-N}{\sqrt{2\pi\sigma_{s2}^2}}\exp\left(-\frac{(s-s_{02})^2}{2\sigma_{s2}^2}\right)
    \label{double-log-normal}
\end{multline}
This function has five parameters. Thus, we have to optimize all five parameters for fitting. The black dashed line in the first panel of Fig.~\ref{Figure 7} is the double-log-normal fit to the numerical PDF at $a=1$. The double-log-normal function fits the numerical PDFs very well, with fit parameters provided in Table~\ref{Table 2}.

The PDFs for the later times ($a=0.1$ and $0.01$) are shown in the second and third row of Figure~\ref{Figure 7}. At these later stages in the contraction of the clouds, the PDFs become close to log-normal. The green dotted lines represent the corresponding log-normal fits via Eq.~(\ref{log-normal PDF}). However, some intermittency and accompanying skewness remains, which is why we also show fits using the \citet{Hopkins2013} intermittency PDF, Eq.~(\ref{Hopkins pdf}) as the blue solid lines (with fit parameters provided in Table~\ref{Table 3}). These provide very good fits to all the simulations at $a=0.1$ and $0.01$, when the clouds have become fairly dense and reached a nearly isothermal, cold state, which can be seen from the temperature projection plots in Fig.~\ref{Figure 5}.

\section{Limitations}
\label{Sec 4}
In this section, we discuss some of the main limitations of our work. As a result of the simplicity of hydrodynamic simulations, comparisons with observational results are limited and should be considered carefully. These limitations are listed below:

\begin{itemize}
\item In this study, we neglect the detailed chemistry of the gas. Throughout the study we consider a mean molecular weight $\mu=1$. However, in reality, $\mu$ changes from about 1.3 in the atomic phase to about 2.3 in the molecular phase. For simplicity, we did not model this change in $\mu$. However, this does not have a significant impact on our general conclusions and would only marginally change our quantitative results. For example, the sound speed would change by a factor $\sim\sqrt{\mu}$, which is significantly less than the differences between our models with different contraction rates.

\item The numerical resolution of our simulations is limited. We have performed all the simulations with a resolution of $512^3$ grid points. However, we provide a resolution study in Appendix~\ref{Sec B}, which demonstrates reasonable convergence of the integrated quantities shown in Fig.~\ref{Figure 2}.

\item In this study, we only consider constant contraction rates (i.e., independent of time or scale factor). While this allows for a simple and clear investigation of the effects of different constant contraction rates, it does not allow us to study the effects of a dynamical change in the contraction rate, which is for example the case for gravitational contraction, where the contraction accelerates over time. Such cases are considered in the previous work by \citet{Robertson_2012}. However, here we wanted to focus on the effects of heating and cooling without specifying the physical source of the contraction (for example, gravity, shock waves, or cloud-cloud collisions) and chose a constant contraction rate for simplicity. Follow-up work may study cases where the contraction rate is time- and scale-dependent.

\end{itemize}

\section{Summary and Conclusions}
\label{Sec 5}
In this paper, we study the compression of magnetized turbulent gas, incorporating the effects of radiative heating and cooling. We investigate whether compression can form molecular clouds from the warm atomic phase, matching observed properties, such as the linewidth--size relation.
We use the grid-based code FLASH for our numerical experiments. The simulations follow the global compression of turbulent gas at moderately initial supersonic velocities (each with a velocity of $11.6\,\text{km/s}$ for HD simulations and $11.1\,\text{km/s}$ for MHD simulations). A total of six simulations were carried out: three different compression rates (Slow, Medium, and Fast), each for HD (no magnetic fields) and MHD (with magnetic fields). In the following we summarise our main results:

\begin{itemize}
\item The global compression enhances the turbulent velocity of purely hydrodynamic turbulence, if the compression timescale ($1/H$) is smaller than the turbulent dissipation timescale ($\tau$). For cases with compression timescale less than dissipation timescale, although the turbulence does not get enhanced, the natural turbulence dissipation gets delayed due to the energy pumping from global compression. However, compared to HD turbulence, the situation in MHD turbulence is slightly different. In the MHD models the magnetic field stores additional energy, which replenishes some of the kinetic energy that is dissipated.
    
\item Initially, when the temperature is high ($\sim 5000\,\text{K}$), the cooling rate is also high and the gas undergoes rapid radiative cooling. For all the simulations, the temperature saturates around 5--$30\,$K when the density has reached $n\sim 10^6\,\text{cm}^{-3}$. 
    
\item When the contraction rate ($|H|$) is high, the cascade of turbulence energy to smaller scales is limited and dissipation becomes inefficient. As a result, the Mach number ($\mathcal{M}$) becomes hypersonic, too large to be compatible with typical values in the Milky Way. On the other hand, when $\omega/|H|\ll 1$, the dissipation dominates and fails to sustain the turbulence. Only if the contraction timescale is of the order of the turbulence dissipation timescale, $\mathcal{M}$ remains in the supersonic regime ($\sim5-10$), consistent with the range of observed Mach numbers in typical molecular clouds in the Milky Way.
    
\item Due to rapid radiative cooling the temperature drops. Thus, the Mach number ($\mathcal{M}$) evolution depends on the balance between turbulence dissipation rate and cooling rate. In the molecular regime, the velocity dispersion for the MHD-Medium simulation shows a strong correlation between velocity dispersion and the size of the cloud ($L$--$\sigma_v$ scaling relation) and falls between the Larson relation and the B08 relation, and almost follows the S87 relation. By contrast, the linewidth--size relations for slow and fast compression do not fit the observed scaling relations.
    
\item It is observationally established that in the high-density regime ($n\gtrsim 10^3\,\text{cm}^{-3}$), the magnetic field strength ($B$) is correlated with the density of the cloud and proportional to $\rho^\kappa$ (for low magnetic field strengths, $\kappa\approx2/3$, and for ambipolar diffusion driven turbulence $\kappa\approx 0.47$). The calculated $B$--$n$ correlation from the medium compression MHD simulations falls between these two scalings, as the plasma-$\beta$, is about 0.1, which means that the magnetic field is dynamically important, also consistent with observations.

\item The relation given by Eq.~\eqref{log density variance HD} between logarithmic density variance ($\sigma_s$) and Mach number ($\mathcal{M}$) derived for isothermal turbulence is found to be consistent with the theoretical prediction in the HD case for a forcing parameter $b\approx0.5$. However, it slightly under-predicts $\sigma_s$ after $a<0.1$ which means for our model of compression the value of $b$ is slightly higher. This can be explained by the fact that in this model, $b$ is not constant, but rather changes with scale factor as we go from solenoidal driving (the initial stage of contraction) to a mixed driving (mixture of solenoidal and compressive components), i.e., $1/3<b<1$. We find that direct computation of the effective driving parameter ($b$) using Eq.~\eqref{b} with $\beta\rightarrow\infty$ gives $b\sim 0.4-0.5$, i.e., the turbulence is a mild mixture of compressive components with the initial solenoidal modes. This is also reasonably consistent with the driving parameter ($b_\chi$) computed by Helmholtz decomposition of the velocity field (Eq.~\ref{driving parameter}). However, for MHD there is a difference between the simulation result and the theoretical prediction (Eq.~\ref{log density variance HD}) when $b\approx0.5$ is assumed. The evolution of $b$ computed from Eq.~\eqref{b} suggests that $b>1$. This discrepancy is because the Alfv\'en Mach number ($\mathcal{M}_A$) drops below 2, which means the magnetic field becomes very strong and the \mbox{$\sigma_s$--$\mathcal{M}$} relation is no longer applicable \citep{Molina}. However, the velocity spectra give reasonable values of the driving parameter ($b_\chi \sim 0.3-0.4$), i.e., the suppression of the compressive modes due to strong magnetic fields leave the turbulence as almost completely solenoidal.

\item Just after the turbulence is fully developed and before the contraction starts, which is when a two-phase medium (WMN and CNM) is established, we find that a double-log-normal density PDF provides a good fit to the simulation data. At later times, during the contraction, almost all of the gas becomes dense and cold, such that the isothermal approximation for the density PDF provides a reasonable approximation, and the simulation PDFs are well described by an intermittency PDF model (c.f.,~Fig.~\ref{Figure 7}).
\end{itemize}

In summary, using idealized simulations, we find that the large-scale compression of the warm, atomic, magnetized ISM can drive turbulence by injecting energy into the system due to compression, if the contraction timescale is less than the turbulence dissipation timescale. The models with contraction timescale similar to dissipation timescale can reasonably produce key observed physical properties of molecular clouds. There are several candidates for causing such turbulent compression: global gravitational contraction on large scales, compression due to stellar feedback (e.g., supernova shock waves), or cloud-cloud collisions.

\section*{Acknowledgements}
We thank Yuval Birnboim and Mark Krumholz for useful discussions about the project at the initial stage. A.~M.~would like to thank the Future Research Talent Travel Award team of The Australian National University for hosting him through the program. C.~F.~acknowledges funding provided by the Australian Research Council (Discovery Project DP170100603 and Future Fellowship FT180100495), and the Australia-Germany Joint Research Cooperation Scheme (UA-DAAD). We further acknowledge high-performance computing resources provided by the Leibniz Rechenzentrum and the Gauss Centre for Supercomputing (grants~pr32lo, pr48pi, and GCS Large-scale project~10391), the Australian National Computational Infrastructure (grant~ek9) in the framework of the National Computational Merit Allocation Scheme and the ANU Merit Allocation Scheme. B.~K.~thanks for funding from the DFG grant BA 3706/15-1. The simulation software FLASH was in part developed by the DOE-supported Flash Center for Computational Science at the University of Chicago.






\appendix
\section{Morphology of HD-flow for medium compression}
\begin{figure*}
  \includegraphics[width=1.0\linewidth]{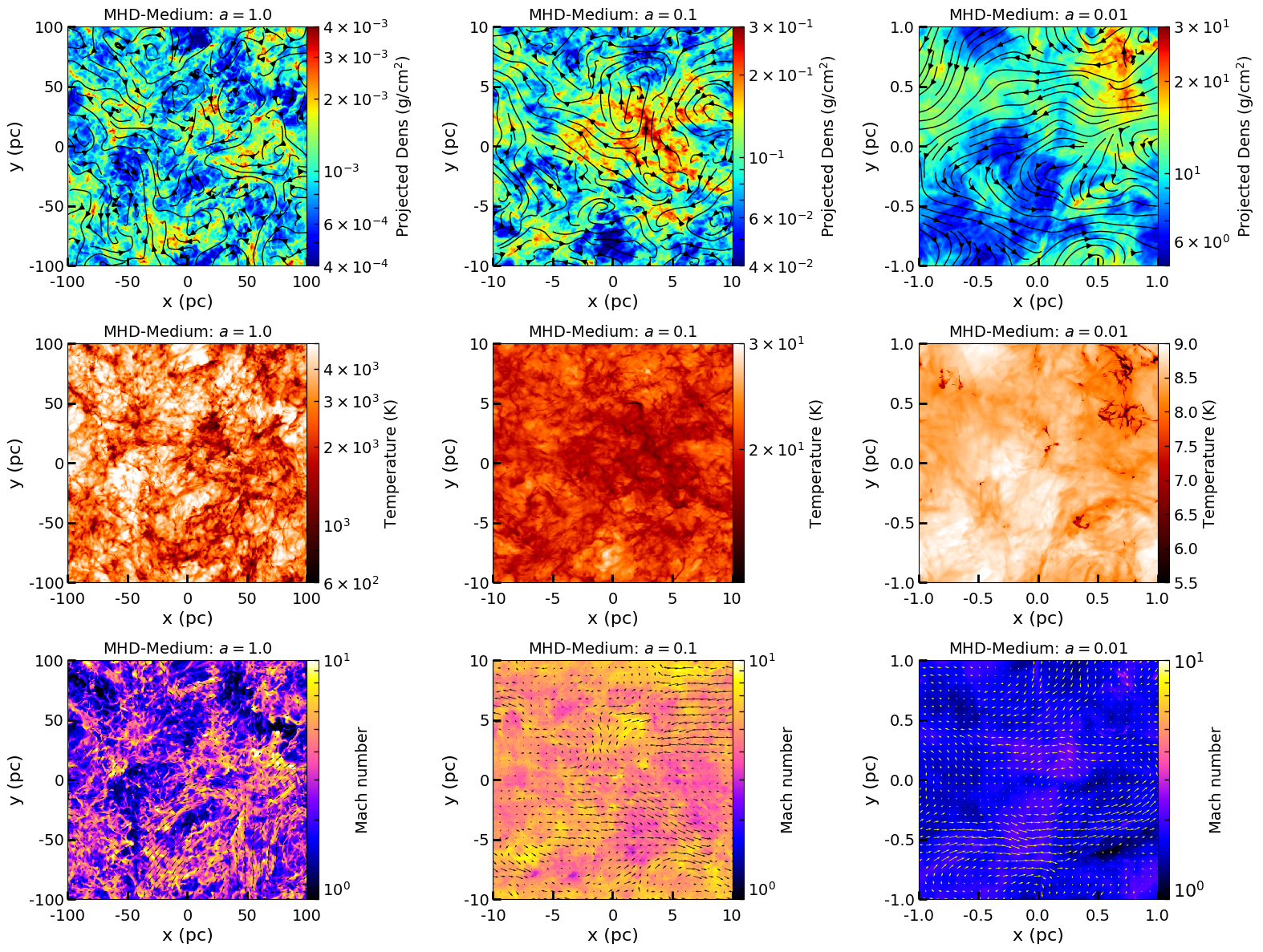}
  \caption{Same as Fig.~\ref{Figure 5}, but for the HD-Medium simulation.}
  \label{Figure A1}
\end{figure*}

Fig.~\ref{Figure A1} shows the flow morphology for the hydrodynamic (HD) simulation with medium compression, similar to the MHD simulation with medium compression shown in Fig.~\ref{Figure 5}.
\FloatBarrier
\section{Convergence with numerical grid resolution}
\label{Sec B}
\begin{figure}
    \centering
    \includegraphics[width=0.99\linewidth]{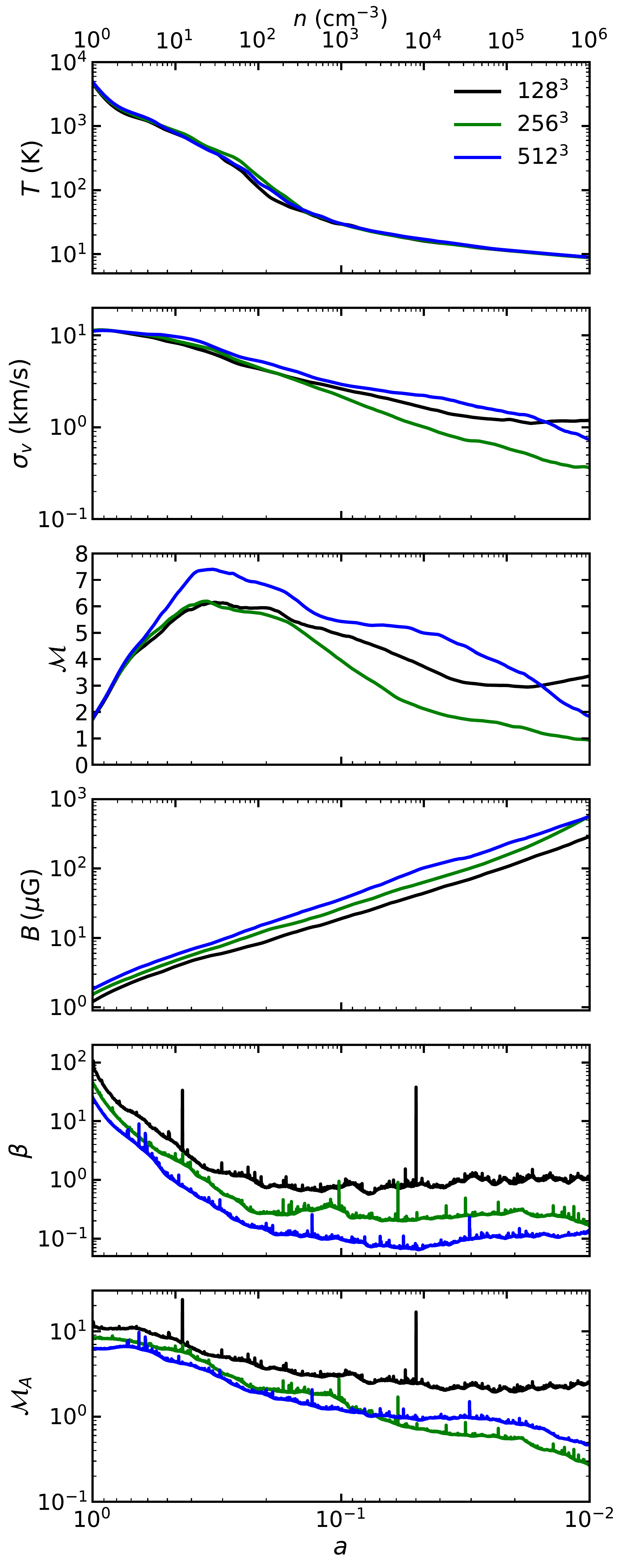}
    \caption{Convergence study of integrated quantities with numerical grid resolution for the MHD-Medium compression ($\omega/H=-1$) model. In each panel, we show results from three different simulations with grid resolutions of $128^3$ (black), $256^3$ (green) and $512^3$ (blue) cells.}
    \label{Figure B1}
\end{figure}

In Fig.~\ref{Figure B1}, we show the dependence and convergence of our numerical results with grid resolution. Fig.~\ref{Figure B1} shows the same integrated quantities as Fig.~\ref{Figure 2} for the MHD-Medium compression ($\omega/H=-1$) model, but at different numerical resolutions. We perform three different simulations with grid resolutions of $128^3$ (black line), $256^3$ (green line) and $512^3$ (blue line). The first panel shows that the variations of temperature with different resolutions are almost negligible when the cloud becomes cold and dense. For the velocity dispersion (second panel of Fig.~\ref{Figure B1}), we see some dependence on numerical resolution, however, there is no clear trend for this dependence. Thus, our end results do not systematically depend on resolution. A similar trend can be seen for the mass-weighted Mach number (third panel of Fig.~\ref{Figure B1}). The quantities that show the strongest resolution dependence involve the magnetic field. As the dynamo amplification and the level of saturation of turbulent dynamo depend on high Reynolds number, which is a resolution-dependent quantity, and the tangling of magnetic field requires higher resolution to resolve, higher grid resolution would produce more converged results. We can clearly see this behaviour in the magnetic field (fourth panel), plasma $\beta$ (fifth panel) and Alfv\'en Mach number (sixth panel).

We note that the fluctuations in plasma $\beta$ and Alfv\'en Mach number are of physical origin. These variables are bursty, because of how they are defined. The bursts come from local fluctuations of thermal and magnetic pressure. Since these can vary substantially on small scales, even the space-averages still show them. For example, if there is just one cell or a small local region that has tiny magnetic pressure, then $\beta$ and $\mathcal{M}_{\mathrm{A}}$ will shoot up (in extreme cases, they might temporarily approach values near infinity, if the magnetic pressure goes to zero in a local region).

Overall, we find that a grid resolution of $128^3$ cells is not enough for simulating the compression of supersonic, magnetized turbulent gas in the context of molecular cloud formation. We need at least a grid resolution of $256^3$ for this study to achieve convergence to within a factor of $\sim2$ at all $a$. We conclude that for our standard resolution of $512^3$ we obtain reasonable, nearly converged results for the integrated cloud quantities that we have focused on in this study.

\bsp    
\label{lastpage}
\end{document}